\documentclass[journal]{IEEEtran}
\usepackage{cite}
\usepackage[dvips]{graphicx}
\usepackage[cmex10]{amsmath}
\usepackage{color}
\usepackage{bm}
\usepackage{theorem}
\usepackage{amscd}
\usepackage{amssymb}

{\theorembodyfont{\normalfont}
\theoremheaderfont{\normalfont\it}
\newtheorem{thm}{ Theorem}
\newtheorem{dfn}[thm]{ Definition}
\newtheorem{lmm}[thm]{ Lemma}

\newtheorem{prp}[thm]{ Proposition}}

{\theorembodyfont{\normalfont}
\theoremheaderfont{\normalfont\it}
\newtheorem{prf}{ Proof:}}

{\theorembodyfont{\normalfont}
\theoremheaderfont{\normalfont\it}
\newtheorem{rmk}{Remark:}}

{\theorembodyfont{\normalfont}
\theoremheaderfont{\normalfont\it}
}

{\theorembodyfont{\normalfont}
\theoremheaderfont{\normalfont\it}
}

{\theorembodyfont{\normalfont}
\theoremheaderfont{\normalfont\it}
}

\newcommand{\bra}[1]{\mbox{$\left\langle#1\right|$}}
\newcommand{\ket}[1]{\mbox{$\left|#1\right\rangle$}}
\newcommand{\inpro}[2]{\mbox{$\left\langle#1|#2\right\rangle$}}

\newcommand{\proj}[1]{\mbox{$\ket{#1}\!\bra{#1}$}}

\begin{document}

\title{The Cost of Randomness for Converting a Tripartite Quantum State to be Approximately Recoverable}

\author{Eyuri Wakakuwa, Akihito Soeda and Mio Murao

\thanks{This work is supported by the Project for Developing Innovation Systems of MEXT, Japan and JSPS KAKENHI (Grant No.~23540463, No.~23240001, No.~26330006, and No.~15H01677).  We also gratefully acknowledge to the ELC project (Grant-in-Aid for Scientific Research on Innovative Areas MEXT KAKENHI (Grant No.~24106009)) for encouraging the research presented in this paper.  This work was presented in part at the 33rd Quantum Information Technology Symposium, 2015, in Japan.}
\thanks{E. Wakakuwa is with the Department of Communication Engineering and Informatics, Graduate School of Informatics and Engineering, The University of Electro-Communications, Japan (email: wakakuwa@quest.is.uec.ac.jp).}
\thanks{A. Soeda is with the Department of Physics, Graduate School of Science, The University of Tokyo, Japan.}
\thanks{M. Murao is with the Department of Physics, Graduate School of Science, The University of Tokyo, and is with Institute for Nano Quantum Information Electronics, The University of Tokyo.}
}

\maketitle

\begin{abstract}
We introduce and analyze a task in which a tripartite quantum state is transformed to an approximately recoverable state by a randomizing operation on one of the three subsystems. We consider cases where the initial state is a tensor product of $n$ copies of a tripartite state $\rho^{ABC}$, and is transformed by a random unitary operation on $A^n$ to another state which is approximately recoverable from its reduced state on $A^nB^n$ (Case 1) or $B^nC^n$ (Case 2). We analyze the minimum cost of randomness per copy required for the task in an asymptotic limit of infinite copies and vanishingly small error of recovery, mainly focusing on the case of pure states. We prove that the minimum cost in Case 1 is equal to the {\it Markovianizing cost} of the state, for which a single-letter formula is known. With an additional requirement on the convergence speed of the recovery error, we prove that the minimum cost in Case 2 is also equal to the Markovianizing cost. Our results have an application for distributed quantum computation.
\end{abstract}

\section{Introduction}

Tripartite quantum states, for which the quantum conditional mutual information (QCMI) is zero, are called {\it quantum Markov chains}, or {\it Markov states} for short\cite{hayden04}. They have been investigated in several contexts, for example, in analyzing the cost of quantum state redistribution \cite{deve08_2,jon09}, investigating effects of the initial system-environment correlation on the dynamics of quantum states \cite{francesco14}, and computing the free energy of quantum many-body systems \cite{poulin11}. A characterization of Markov states is obtained in \cite{hayden04}, in which the following three properties are proved to be equivalent:
\begin{enumerate}
\item {\it Vanishing QCMI}: A tripartite quantum state $\rho^{ABC}$ satisfies
\begin{eqnarray}
I(A:C|B)_\rho=0.\label{eq:markovianity}
\end{eqnarray}
\item {\it Recoverability}: $\rho^{ABC}$ is recoverable from its bipartite reduced state on $AB$ and $BC$, that is, there exist quantum operations ${\mathcal R}:B\rightarrow AB$ and ${\mathcal R}':B\rightarrow BC$ such that 
\begin{eqnarray}
\rho^{ABC}={\mathcal R}(\rho^{BC})={\mathcal R}'(\rho^{AB}).\label{eq:recoverability}
\end{eqnarray}
\item {\it Decomposability}: $\rho^{ABC}$ is equivalent to the following state up to a local unitary transformation on $B$:
\begin{eqnarray}
\sum_jp_j\proj{j}^{b_0}\otimes\sigma_j^{Ab_L}\otimes\phi_j^{b_RC}.\label{eq:decomposability}
\end{eqnarray} 
\end{enumerate}

The equivalence among the three properties, however, breaks down when we require that Equalities (\ref{eq:markovianity}), (\ref{eq:recoverability}) and (\ref{eq:decomposability}) hold {\it approximately}, instead of requiring {\it exactly}. On the one hand, the result by Fawzi and Renner\cite{fawzi15} proves that a tripartite state is recoverable with a small error (i.e., approximately recoverable) if QCMI of the state is small (see also \cite{brandao14,sesh14,berta15,wilde15,berta15_2}). On the other hand, QCMI of a state can be vanishingly small, even if the state does not fit into any decomposition in the form of (\ref{eq:decomposability}) unless significantly deformed (i.e., even if the state is {\it not} approximately decomposable)\cite{ibinson08}. Although the difference in the choices of the distance measures should be carefully taken into account, one could argue that the two results show inequivalence between approximate recoverability and approximate decomposability. This is in contrast to the classical case, for which the corresponding properties are equivalent.

The purpose of this paper is to investigate the relation between approximate recoverability and approximate decomposability from an information theoretical point of view. To this end, we introduce and analyze two information theoretical tasks: {\it Markovianization in terms of recoverability (M-Rec)}, and {\it Markovianization in terms of decomposability (M-Dec).} In both tasks, a tensor product of $n$ copies of a tripartite state $\rho^{ABC}$ is transformed by a random unitary operation on $A^n$, the $n$-copy system of $A$. In the former, the state after the transformation is required to be recoverable up to a small error $\epsilon$. In the latter, the state is supposed to fit into a decomposition of $B^n$ into three subsystems ${\hat b}_0$, ${\hat b}_L$ and ${\hat b}_R$ as (\ref{eq:decomposability}), up to a small error $\epsilon$. We analyze and compare the minimum cost of randomness per copy required for each task, by considering an asymptotic limit of $\epsilon\rightarrow0$ and $n\rightarrow\infty$.

Depending on the type of the recovery map to be applied, we consider two cases for M-Rec. In the first case, the state after the transformation is required to be approximately recoverable from its reduced state on $B^nC^n$, whereas in the second case it is supposed to be approximately recoverable from the reduced state on $A^nB^n$. We call the minimum cost of randomness in each case as the {\it Markovianizing cost in terms of recoverability (M-Rec cost)}, and denote it by $M_{A|BC}^R(\rho)$ and $M_{A|AB}^R(\rho)$, respectively for each case above.

In the previous work\cite{waka15_markov_paper}, we introduced a similar task that we simply call {\it Markovianization}, in which $n$ copies of a tripartite state $\rho^{ABC}$ is transformed by a random unitary operation on $A^n$ to another state which is $\epsilon$-close to a Markov state conditioned by $B^n$. As we prove later, this version of Markovianization is equivalent to M-Dec, up to a dimension-independent rescaling of $\epsilon$. Consequently, the minimum cost of randomness per copy required for Markovianization in the version of \cite{waka15_markov_paper} is equal to the one required for M-Dec. We call the latter as the {\it Markovianizing cost in terms of decomposability (M-Dec cost)}, and denote it by $M_{A|B}^D(\rho)$. A single-letter formula for the M-Dec cost of pure states is hence equal to the one obtained in \cite{waka15_markov_paper}.  

In this paper, we mainly focus on cases where the initial state is pure, that is, $\rho^{ABC}=|\Psi\rangle\!\langle\Psi|$.
The main results of this paper are as follows. First, we prove that $M_{A|BC}^R(\Psi)=M_{A|B}^D(\Psi)$ holds. Second, we prove that $M_{A|AB}^R(\Psi)=M_{A|B}^D(\Psi)$ holds as well, under an additional requirement that the error of recovery converges to zero faster than $1/n$. Thereby we reveal that the gap between approximate recoverability and approximate decomposability disappears in this information theoretical framework, at least in the case of pure states. The obtained results are applied to an analysis of distributed quantum computation in \cite{waka15_2}.

The structure of this paper is as follows. In Section \ref{sec:preliminaries}, we introduce rigorous definitions of approximate recoverability and approximate decomposability, and analyze relations among these conditions. In Section \ref{sec:results}, we introduce the formal definitions of Markovianization in terms of recoverability and that in terms of decomposability, and describe the main results. In Section \ref{sec:alternative}, we introduce and analyze an extension of Markovianization into the one induced by a measurement. Conclusions are given in Section \ref{sec:conclusion}. See Appendices for proofs of the main theorems.

{\it Notations.}  We follow the notations introduced in \cite{waka15_markov_paper}.

\section{Recoverability and Decomposability}\label{sec:preliminaries}

In this section, we present rigorous definitions of approximate recoverability and approximate decomposability. We then prove some general relations among these.

Let us first present three equivalent definitions for ``exact Markovness'' of tripartite quantum states.

\begin{thm}\label{thm:markovdechayden}(Theorem 6 in \cite{hayden04})
The following three conditions are equivalent:
\begin{enumerate}
\item $\Upsilon^{ABC}$ satisfies $I(A:C|B)_\Upsilon=0$.
\item There exist quantum operations ${\mathcal R}:B\rightarrow AB$ and ${\mathcal R}':B\rightarrow BC$ such that 
\begin{eqnarray}
\Upsilon^{ABC}={\mathcal R}(\Upsilon^{BC})={\mathcal R}'(\Upsilon^{AB}).\nonumber
\end{eqnarray}
\item There exist three Hilbert spaces ${\mathcal H}^{b_{\scalebox{0.45}{$0$}}}$, ${\mathcal H}^{b_{\scalebox{0.45}{$L$}}}$, ${\mathcal H}^{b_{\scalebox{0.45}{$R$}}}$ and an isometry $\Gamma:{\mathcal H}^B_{\Upsilon}\rightarrow{\mathcal H}^{b_{\scalebox{0.45}{$0$}}}\otimes{\mathcal H}^{b_{\scalebox{0.45}{$L$}}}\otimes{\mathcal H}^{b_{\scalebox{0.45}{$R$}}}$ such that $\Upsilon^{ABC}$ is decomposed as
\begin{eqnarray}
\Gamma^B\Upsilon^{ABC}\Gamma^{\dagger B}=\sum_{i}q_i\proj{i}^{b_{\scalebox{0.45}{$0$}}}\otimes\sigma_i^{Ab_{\scalebox{0.45}{$L$}}}\otimes\phi_i^{b_{\scalebox{0.45}{$R$}}C}
\label{eq:markovdecofstate}
\end{eqnarray}
with some probability distribution $\{q_i\}_{i}$, orthonormal basis $\{\ket{i}\}_i$ of ${\mathcal H}^{b_{\scalebox{0.45}{$0$}}}$, states $\sigma_i\in{\mathcal S}({\mathcal H}^A\otimes{\mathcal H}^{b_{\scalebox{0.45}{$L$}}})$ and $\phi_i\in{\mathcal S}({\mathcal H}^{b_{\scalebox{0.45}{$R$}}}\otimes{\mathcal H}^{C})$.
\end{enumerate}
\end{thm}
A tripartite quantum state that satisfies the conditions in the above theorem is called a {\it Markov state conditioned by $B$}. When $\Upsilon^{ABC}$ is a Markov state conditioned by $B$, (\ref{eq:markovdecofstate}) is called a {\it Markov decomposition of $\Upsilon^{ABC}$}, and $\Gamma$ in (\ref{eq:markovdecofstate}) is called a {\it Markov isometry on $B$ with respect to $\Upsilon^{ABC}$}.

We now introduce four different characterizations of a tripartite quantum state being ``approximately Markov''.

\begin{dfn}\label{df:eqcmi}
A tripartite state $\rho^{ABC}$ is {\it $\epsilon$-QCMI conditioned by $B$} if it satisfies
\begin{eqnarray}
I(A:C|B)_\rho\leq\epsilon.\nonumber
\end{eqnarray}
\end{dfn}

\begin{dfn}
A tripartite state $\rho^{ABC}$ is {\it $\epsilon$-recoverable from $BC$} if there exists a quantum operation ${\mathcal R}:B\rightarrow AB$ such that
\begin{eqnarray}
\left\|\rho^{ABC}-{\mathcal R}(\rho^{BC})\right\|_1\leq\epsilon.\nonumber
\end{eqnarray}
\end{dfn}

\begin{dfn}\label{dfn:errec}
A tripartite state $\rho^{ABC}$ is {\it $\epsilon$-recoverable from $AB$} if there exists a quantum operation ${\mathcal R}':B\rightarrow BC$ such that
\begin{eqnarray}
\left\|\rho^{ABC}-{\mathcal R}'(\rho^{AB})\right\|_1\leq\epsilon.\nonumber
\end{eqnarray}
\end{dfn}

\begin{dfn}\label{dfn:emarkov}
A tripartite state $\rho^{ABC}$ is {\it $\epsilon$-decomposable on $B$} if there exists a Markov state $\Upsilon^{ABC}$ conditioned by $B$ such that
\begin{eqnarray}
\left\|\rho^{ABC}-\Upsilon^{ABC}\right\|_1\leq\epsilon.\label{eq:appmarkov}
\end{eqnarray}
\end{dfn}

As we prove in Appendix \ref{app:markdec}, Condition (\ref{eq:appmarkov}) is equivalent to the condition that $\rho^{ABC}$ fits into the best possible choice of the tensor-product decomposition of $B$ into three subsystems as (\ref{eq:decomposability}), up to a small error $\epsilon$. This fact supports the use of ``decomposable'' in Definition \ref{dfn:emarkov}. 

The following relations hold among the conditions described above.

\begin{lmm}\label{lmm:markovrecoverable}
For an arbitrary tripartite state $\rho^{ABC}$:
\begin{enumerate}
\item $\rho^{ABC}$ is $2\sqrt{\ln{2}}\sqrt{\epsilon}$-recoverable from $AB$ and $BC$ if it is $\epsilon$-QCMI conditioned by $B$.
\item $\rho^{ABC}$ is $f(\epsilon,d_C)$-recoverable from $AB$ if it is $\epsilon$-recoverable from $BC$ and $\epsilon\leq1$. Here $f(\epsilon,d):=2\sqrt{\ln{2}}\sqrt{4\epsilon\log{d}+2h(\epsilon)}$ and $h(\epsilon)$ is the binary entropy defined by $h(\epsilon):=-\epsilon\log{\epsilon}-(1-\epsilon)\log{(1-\epsilon)}$.
\item $\rho^{ABC}$ is $2\epsilon$-recoverable from $AB$ and $BC$ if it is $\epsilon$-decomposable on $B$.
\end{enumerate}
\end{lmm}

\begin{prf}

Property 1) is proved in \cite{fawzi15} (see Inequality (6) therein). As for Property 2), suppose $\rho^{ABC}$ is $\epsilon$-recoverable from $BC$. There exists a linear CPTP map ${\mathcal R}:B\rightarrow AB$ such that
\begin{eqnarray}
\left\|\rho^{ABC}-{\mathcal R}(\rho^{BC})\right\|_1\leq\epsilon.\nonumber
\end{eqnarray}
Due to Inequality (8.28) in \cite{berta15}, we have
\begin{eqnarray}
I(A:C|B)_\rho\leq4\epsilon\log{d_C}+2h(\epsilon).\nonumber
\end{eqnarray}
Applying Property 1), we obtain 2).

Property 3) is proved as follows. Suppose $\rho^{ABC}$ is $\epsilon$-decomposable on $B$, and let $\Upsilon^{ABC}$ be a Markov state conditioned by $B$ satisfying (\ref{eq:appmarkov}). There exist quantum operations ${\mathcal R}:B\rightarrow AB$ and ${\mathcal R}':B\rightarrow BC$ such that $\Upsilon^{ABC}={\mathcal R}(\Upsilon^{BC})={\mathcal R}'(\Upsilon^{AB})$. From (\ref{eq:appmarkov}) and the monotonicity of the trace distance, we have
\begin{eqnarray}
\left\|{\mathcal R}(\rho^{BC})-\Upsilon^{ABC}\right\|_1\leq\epsilon,\;\;\left\|{\mathcal R}'(\rho^{AB})-\Upsilon^{ABC}\right\|_1\leq\epsilon.\nonumber
\end{eqnarray}
By the triangle inequality, we obtain
\begin{eqnarray}
\left\|\rho^{ABC}-{\mathcal R}(\rho^{BC})\right\|_1\leq2\epsilon,\;\;\left\|\rho^{ABC}-{\mathcal R}'(\rho^{AB})\right\|_1\leq2\epsilon,\nonumber
\end{eqnarray}
which completes the proof of Property 3).
\hfill$\blacksquare$
\end{prf}

\section{Markovianizing Costs}
\label{sec:results}

In this section, we present a concept of Markovianization, and describe the main results on the Markovianizing costs of tripartite quantum states. Proofs are given in Appendix \ref{prf:leftrightmark} and \ref{prf:leftmarkpure}.

Let us first present Markovianization as formulated in \cite{waka15_markov_paper}.

\begin{dfn}\label{dfn:mofpsi}(Equivalent to Definition 7 in \cite{waka15_markov_paper})
A tripartite state $\rho^{ABC}$ is {\it Markovianized with the randomness cost $R$ on $A$, conditioned by $B$}, if the following statement holds. That is, there exists a sequence of sets of unitaries $\{\{V_{n,k}\}_{k=1}^{2^{nR}}\}_{n=1}^\infty$, with each $V_{n,k}$ acting on $({\mathcal H}^A)^{\otimes n}$, such that ${\mathcal V}_n((\rho^{ABC})^{\otimes n})$ is $\epsilon_n$-decomposable on ${\bar B}$ for ${\mathcal V}_n:\tau\mapsto2^{-nR}\sum_{k=1}^{2^{nR}}V_{n,k}\tau V_{n,k}^{\dagger}$ and $\lim_{n\rightarrow\infty}\epsilon_n=0$.

The {\it Markovianizing cost} of $\rho^{ABC}$ is defined as  $M_{A|B}(\rho^{ABC}):=\inf\{R\:|\:\rho^{ABC}$ is Markovianized with the randomness cost $R$ on $A$, conditioned by $B\}$.
\end{dfn}
We refer to the Markovianization of Definition \ref{dfn:mofpsi} as the {\it Markovianization in terms of decomposability (M-Dec)} in the rest. Correspondingly, we call $M_{A|B}(\rho^{ABC})$ as the Markovianizing cost in terms of decomposability (M-Dec cost), and denote it by $M_{A|B}^D(\rho^{ABC})$. A single-letter formula for the M-Dec cost of tripartite pure states is obtained in \cite{waka15_markov_paper} (See Appendix \ref{app:marksingle}).

Let us now introduce the idea of the Markovianizing cost in terms of recoverability (M-Rec cost). Depending on the type of the recovery map to be applied, we have two different formulations for the M-Rec cost.

\begin{dfn}
A tripartite state $\rho^{ABC}$ is {\it Markovianized with the randomness cost $R$ on $A$, in terms of recoverability from $BC$}, if the following statement holds. That is, there exists a sequence of sets of unitaries $\{\{V_{n,k}\}_{k=1}^{2^{nR}}\}_{n=1}^\infty$, with each $V_{n,k}$ acting on $({\mathcal H}^A)^{\otimes n}$, such that ${\mathcal V}_n((\rho^{ABC})^{\otimes n})$ is $\epsilon_n$-recorerable from ${\bar B}{\bar C}$ for ${\mathcal V}_n:\tau\mapsto2^{-nR}\sum_{k=1}^{2^{nR}}V_{n,k}\tau V_{n,k}^{\dagger}$ and $\lim_{n\rightarrow\infty}\epsilon_n=0$.

The {\it Markovianizing cost of $\rho^{ABC}$ in terms of recoverability from $BC$} is defined as  $M^R_{A|BC}(\rho^{ABC}):=\inf\{R\:|\:\rho^{ABC}$ is Markovianized with the randomness cost $R$ on $A$, in terms of recoverability from $BC\}$.
\end{dfn}

\begin{dfn}\label{dfn:rmarkovi}
A tripartite state $\rho^{ABC}$ is {\it Markovianized with the randomness cost $R$ on $A$, in terms of recoverability from $AB$}, if the following statement holds. That is, there exists a sequence of sets of unitaries $\{\{V_{n,k}\}_{k=1}^{2^{nR}}\}_{n=1}^\infty$, with each $V_{n,k}$ acting on $({\mathcal H}^A)^{\otimes n}$, such that ${\mathcal V}_n((\rho^{ABC})^{\otimes n})$ is $\epsilon_n$-recorerable from ${\bar A}{\bar B}$ for ${\mathcal V}_n:\tau\mapsto2^{-nR}\sum_{k=1}^{2^{nR}}V_{n,k}\tau V_{n,k}^{\dagger}$ and $\lim_{n\rightarrow\infty}\epsilon_n=0$.

The {\it Markovianizing cost of $\rho^{ABC}$ in terms of recoverability from $AB$} is defined as  $M^R_{A|AB}(\rho^{ABC}):=\inf\{R\:|\:\rho^{ABC}$ is Markovianized with the randomness cost $R$ on $A$, in terms of recoverability from $AB\}$.
\end{dfn}

The following two theorems are the main results of this paper. The first one (Theorem \ref{thm:leftrightmark}) shows general properties of the M-Rec costs of an arbitrary (possibly mixed) tripartite state, and the second one (Theorem \ref{thm:leftmarkpure}) states that the three types of the Markovianizing cost are equal for pure states. We also present a lemma that plays a central role in the proof of Theorem \ref{thm:leftmarkpure}. Proofs are given in Appendix \ref{prf:leftrightmark} and \ref{prf:leftmarkpure}.

\begin{thm}\label{thm:leftrightmark}
For any tripartite state $\rho^{ABC}$, we have
\begin{eqnarray}
I(A:C|B)_\rho\leq M^R_{A|BC}(\rho^{ABC})\leq M_{A|B}^D(\rho^{ABC})\label{eq:leftmark}
\end{eqnarray}
and
\begin{eqnarray}
I(A:C|B)_\rho\leq M^R_{A|AB}(\rho^{ABC})\leq M_{A|B}^D(\rho^{ABC}).\label{eq:rightmark}
\end{eqnarray}
\end{thm}

\begin{thm}\label{thm:leftmarkpure}
For any tripartite pure state $\Psi^{ABC}$, we have
\begin{eqnarray}
M^R_{A|BC}(\Psi^{ABC})=M_{A|B}^D(\Psi^{ABC}).\label{eq:leftmarkpure}
\end{eqnarray}
If we additionally require in Definition \ref{dfn:rmarkovi} that
\begin{eqnarray}
\lim_{n\rightarrow\infty}n\cdot\epsilon_n=0,\label{eq:additional}
\end{eqnarray}
we also have
\begin{eqnarray}
M^R_{A|AB}(\Psi^{ABC})=M_{A|B}^D(\Psi^{ABC}).\label{eq:rightmarkpure}
\end{eqnarray}
\end{thm}

\begin{lmm}\label{thm:corrtran}
Let $|\Psi^{ABC}\rangle$ be a pure state, and for any $n$ and $\epsilon\in(0,1)$, let $\mathcal E$ be a quantum operation on ${\bar A}$ that satisfy
\begin{eqnarray}
\left\|({\mathcal E}^{{\bar A}}\!\!\otimes{\rm id}^{{\bar C}})\left((\Psi^{AC})^{\otimes n}\right)-(\Psi^{AC})^{\otimes n}\right\|_1\leq\epsilon.\label{eq:cortranerror}
\end{eqnarray}
Then we have
\begin{eqnarray}
&&\frac{1}{n}I({\bar A}:{\bar B}{\bar C})_{{\mathcal E}^{{\bar A}}(\Psi^{\otimes n})}\nonumber\\
&&\geq M_{A|B}^D(\Psi^{ABC})-5\eta(\zeta_{{}_\Psi}\!(\epsilon))\log{d_A}.\nonumber
\end{eqnarray}
Here, $\zeta_{{}_\Psi}\!(\epsilon)$ is a function that satisfies $\lim_{\epsilon\rightarrow0}\zeta_{{}_\Psi}\!(\epsilon)=0$, and does not depend on $n$. $\eta(x)$ is a function defined by
\begin{eqnarray}
\eta(x):=\begin{cases}
x-x\log{x}&(x\leq 1/e)\\
x+\frac{1}{e}&(x\geq 1/e)
\end{cases},\nonumber
\end{eqnarray}
where $e$ is the base of natural logarithm. 
\end{lmm}

It is left open whether Equality (\ref{eq:rightmarkpure}) holds when we drop Condition (\ref{eq:additional}). An underlying problem is whether we can eliminate the dimension dependence of the error in Property 2) in Lemma \ref{lmm:markovrecoverable}. We formulate this problem by the following proposition.
\begin{prp}\label{prp:symrec}(unproven) 
There exists a nonnegative function $g(\epsilon)$, which is independent of dimensions of quantum systems and satisfies $\lim_{\epsilon\rightarrow0}g(\epsilon)=0$, such that the following statement holds for an arbitrary tripartite state $\rho^{ABC}$ and $\epsilon>0$: The state $\rho^{ABC}$ is $g(\epsilon)$-recoverable from $BC$ if it is $\epsilon$-recoverable from $AB$.
\end{prp}
Condition (\ref{eq:additional}) in Theorem \ref{thm:leftmarkpure} can be eliminated if the above proposition is true. The reason is as follows: If ${\mathcal V}_n((\rho^{ABC})^{\otimes n})$ is $\epsilon$-recoverable from ${\bar A}{\bar B}$, the state is $g(\epsilon)$-recoverable from ${\bar B}{\bar C}$. Thus we have $M^R_{A|BC}(\Psi^{ABC})\leq M^R_{A|AB}(\Psi^{ABC})$, which implies Equality (\ref{eq:rightmarkpure}) when combined with (\ref{eq:rightmark}) and (\ref{eq:leftmarkpure}).

\section{Alternative Formulations}\label{sec:alternative}

\begin{figure}[t]
\begin{center}
\includegraphics[bb={0 0 722 348}, scale=0.33]{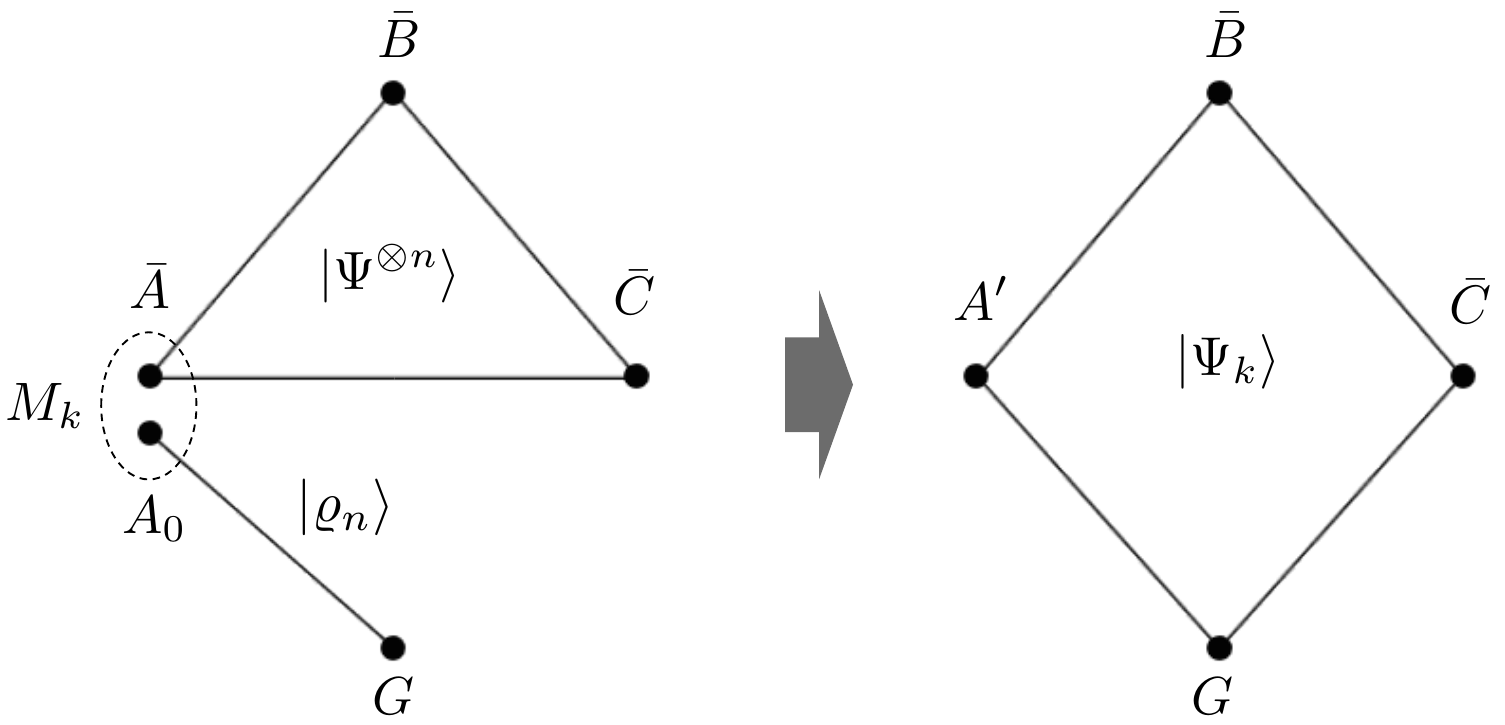}
\end{center}\caption{A graphical representation of measurement-induced Markovianization of a pure state. After the measurement, the reduced state on $A'{\bar B}{\bar C}$ is required to be approximately recoverable.}
\label{fig:markovmeasure}
\end{figure}

In this section, we introduce an extension of M-Rec to that by a measurement (Figure \ref{fig:markovmeasure}), which will be referred to as {\it measurement-induced Markovianization in terms of recoverability}. In particular, we consider an extension of the M-Rec cost in Definition \ref{dfn:rmarkovi} to that by a measurement. The result obtained here has a direct application for distributed quantum computation\cite{waka15_2}.

Let $|\Psi\rangle^{ABC}$ be a tripartite pure state, and let $|\varrho\rangle^{A_0G}$ be a bipartite pure entangled state shared by Alice and George. Consider a state transformation of $(\Psi^{\otimes n})^{{\bar A}{\bar B}{\bar C}}\otimes\proj{\varrho}^{A_0G}$ induced by a measurement on ${\bar A}A_0$, which is described by a set of measurement operators $\{M^{{\bar A}A_0\rightarrow A'}_{ k}\}_{k}$. The probability of obtaining the measurement outcome $k$ is given by
\begin{eqnarray}
p_{ k}=\|M_{ k}|\Psi^{\otimes n}\rangle^{{\bar A}{\bar B}{\bar C}}|{\varrho}\rangle^{A_0G}\|^2,\nonumber
\end{eqnarray}
and the post-measurement state corresponding to the outcome $k$ is given by
\begin{eqnarray}
|\Psi_{ k}\rangle^{A'{\bar B}{\bar C}G}=\frac{1}{\sqrt{p_{ k}}}M_{ k}|\Psi^{\otimes n}\rangle^{{\bar A}{\bar B}{\bar C}}|{\varrho}\rangle^{A_0G}.\label{eq:postmeask}
\end{eqnarray}
We require that (i) the measurement does not significantly change the reduced state on ${\bar B}{\bar C}$ on average, and (ii) the reduced state of the post-measurement state (\ref{eq:postmeask}) on $A'{\bar B}{\bar C}$ is approximately recoverable from $A'{\bar B}$ on average. We focus on the minimum amount of a correlation between systems ${\bar B}{\bar C}$ and $G$, which is inevitably generated by the measurements that satisfy the two conditions. A precise definition is given as follows. 

\begin{dfn}\label{dfn:rofpsi}
A pair $(\ket{{\varrho}}^{A_0G},\{M^{{\bar A}A_0\rightarrow A'}_{ k}\}_k)$ is called {\it an $(n,R,\epsilon)$-Markovianization pair for a tripartite pure state $|\Psi\rangle^{ABC}$} if it satisfies the following conditions:
\begin{enumerate}
\item The measurement does not significantly change the reduced state on ${\bar B}{\bar C}$ on average, that is,
\begin{eqnarray}
\sum_{k}p_{ k}\left\|(\Psi^{\otimes n})^{{\bar B}{\bar C}}-\Psi^{{\bar B}{\bar C}}_{ k}\right\|_1\leq{\epsilon}.\label{eq:recoverabilite2}
\end{eqnarray}
\item There exist quantum operations ${\mathcal R}_{ k}:{\bar B}\rightarrow {\bar B}{\bar C}$ that satisfy
\begin{eqnarray}
\sum_{k}p_{ k}\left\|\Psi^{A'{\bar B}{\bar C}}_{ k}-{\mathcal R}_{ k}(\Psi^{A'{\bar B}}_{ k})\right\|_1\leq{\epsilon}.\label{eq:recoverabilize}
\end{eqnarray}
\item The correlation between ${\bar B}{\bar C}$ and $G$ produced by the measurement is at most $nR$ bits in QMI on average, that is,
\begin{eqnarray}
I(G:{\bar B}{\bar C})_{av}:=\sum_{k}p_{ k}I(G:{\bar B}{\bar C})_{\Psi_{ k}}\leq nR.\label{eq:iavebou}
\end{eqnarray}
\end{enumerate}  
A state $|\Psi\rangle^{ABC}$ is {\it Markovianized with the correlation production $R$ by a measurement on $A$, in terms of recoverability from $AB$}, if there exists a sequence of $(n,R,\epsilon_n)$-Markovianization pair ($n=1,2,\cdots$) such that $\lim_{n\rightarrow\infty}\epsilon_n=0$.

The {\it measurement-induced Markovianizing cost of $|\Psi\rangle^{ABC}$ in terms of recoverability from $AB$} is defined as $M^{R,m}_{A|AB}(\Psi^{ABC}):=\inf\{R\;|\;|\Psi\rangle^{ABC}$ is {\it Markovianized} with the correlation production $R$ by a measurement on $A$, in terms of recoverability from $AB\}$.
\end{dfn}

The measurement-induced Markovianizing cost of pure states defined as above is equal to the Markovianizing cost in terms of random unitary operations, as presented by the following theorem. A proof is given in Appendix \ref{app:measmark}.

\begin{thm}\label{thm:rightmark}
For any tripartite pure state $\Psi^{ABC}$, we have
\begin{eqnarray}
M^{R,m}_{A|AB}(\Psi^{ABC})=M^D_{A|B}(\Psi^{ABC}),\label{eq:mrmeqmd}
\end{eqnarray}
if we additionally require in Definition \ref{dfn:rofpsi} that
\begin{align}
\lim_{n\rightarrow\infty}n\cdot \epsilon_n=0.\label{eq:additionall}
\end{align}
This additional condition can be eliminated if Proposition \ref{prp:symrec} is true.
\end{thm}

\section{Conclusions and Discussions}\label{sec:conclusion}

We have introduced the task of Markovianization in terms of recoverability (M-Rec), and that in terms of decomposability (M-Dec). The latter of which turns out to be equivalent to Markovianization in the version of our previous paper \cite{waka15_markov_paper}. For pure states, we have proven that the minimum cost of randomness required for M-Rec is equal to the one required for M-Dec, for which a single-letter formula has been known. Our results have applications in analyzing optimal costs of resources in distributed quantum computation\cite{waka15_2}. An open question is whether Equalities (\ref{eq:rightmarkpure}) and (\ref{eq:mrmeqmd}) holds when we drop Condition (\ref{eq:additional}). Another related question is whether we can eliminate the dimension dependence of the error in Property 2) in Lemma \ref{lmm:markovrecoverable}.

\hfill

\noindent{\it Note added:} After the completion of this work, the authors have been informed about another work \cite{wildenote}, in which a task similar to M-Rec in our paper was independently proposed. Their definition of the task is more general than ours, in that they consider ``coordinated'' random unitary operations over systems ${\bar A}$, ${\bar B}$ and ${\bar C}$ for Markovianizing a state. They independently derived a lower bound on the cost of randomness, from which the first inequalities in (\ref{eq:leftmark}) and (\ref{eq:rightmark}) are derived as a corollary.

\section*{Acknowledgments}

The authors thank Mario Berta and Tomohiro Ogawa for helpful discussions. EW is grateful for Mark Wilde informing the authors about Ref. \cite{wildenote} and Inequality (8.28) of \cite{berta15}, regarding Property 2) in Lemma \ref{lmm:markovrecoverable}.

\bibliographystyle{IEEEtran}
\bibliography{markov}

%
%

\appendices

\section{Approximate Decomposability}\label{app:markdec}

In this appendix, we prove that there exists a Markov state $\Upsilon^{ABC}$ satisfying Inequality (\ref{eq:appmarkov}), if and only if $\rho^{ABC}$ is $\epsilon$-invariant when ``squeezed'' into a decomposition of $B$ in the form of (\ref{eq:decomposability}), up to a dimension-independent rescaling of $\epsilon$. Thereby we justify referring to Condition (\ref{eq:appmarkov}) as ``$\epsilon$-decomposability'', and to Markovianization in the version of Definition \ref{dfn:mofpsi} as ``Markovianization in terms of decomposability''.

Consider three Hilbert spaces ${\mathcal H}^{b_{\scalebox{0.45}{$0$}}}$, ${\mathcal H}^{b_{\scalebox{0.45}{$L$}}}$, ${\mathcal H}^{b_{\scalebox{0.45}{$R$}}}$ and a linear isometry $\Gamma:{{\mathcal H}^B}\rightarrow{\mathcal H}^{b_{\scalebox{0.45}{$0$}}}\otimes{\mathcal H}^{b_{\scalebox{0.45}{$L$}}}\otimes{\mathcal H}^{b_{\scalebox{0.45}{$R$}}}$ such that
\begin{align}
{\rm img}\Gamma=\bigoplus_i{\mathcal H}_i^{b_{\scalebox{0.45}{$0$}}}\otimes {\mathcal H}_i^{b_{\scalebox{0.45}{$L$}}}\otimes {\mathcal H}_i^{b_{\scalebox{0.45}{$R$}}}.\label{eq:gamma2dagger}
\end{align}
Here, ${\mathcal H}_i^{b_{\scalebox{0.45}{$0$}}}$ are one-dimensional subspaces of ${\mathcal H}^{b_{\scalebox{0.45}{$0$}}}$ spanned by $|i\rangle$, with $\{|i\rangle\}_i$ being an orthonormal basis of ${\mathcal H}^{b_{\scalebox{0.45}{$0$}}}$. ${\mathcal H}_i^{b_{\scalebox{0.45}{$L$}}}$ and ${\mathcal H}_i^{b_{\scalebox{0.45}{$R$}}}$ are subspaces of ${\mathcal H}^{b_{\scalebox{0.45}{$L$}}}$ and ${\mathcal H}^{b_{\scalebox{0.45}{$R$}}}$, respectively. 
Define a map $T_{\Gamma,\{|i\rangle\}}$ on ${\mathcal S}({\mathcal H}^A\otimes{\mathcal H}^B\otimes{\mathcal H}^C)$ by
\begin{eqnarray}
T_{\Gamma,\{|i\rangle\}}({\rho}^{ABC})=\Gamma^\dagger\left(\sum_ip_i\proj{i}^{b_{\scalebox{0.45}{$0$}}}\otimes\rho_i^{Ab_{\scalebox{0.45}{$L$}}}\otimes\rho_i^{b_{\scalebox{0.45}{$R$}}C}\right)\Gamma,\label{eq:markovdecrho}
\end{eqnarray}
where
\begin{eqnarray}
p_i&=&{\rm Tr}\left[\langle i|^{b_{\scalebox{0.45}{$0$}}}\Gamma\rho\Gamma^\dagger|i\rangle^{b_{\scalebox{0.45}{$0$}}}\right],\nonumber\\
\rho_i^{Ab_{\scalebox{0.45}{$L$}}}&=&p_i^{-1}{\rm Tr}_{b_{\scalebox{0.45}{$R$}}C}\left[\langle i|^{b_{\scalebox{0.45}{$0$}}}\Gamma\rho\Gamma^\dagger|i\rangle^{b_{\scalebox{0.45}{$0$}}}\right]\nonumber\
\end{eqnarray}
and
\begin{eqnarray}
\rho_i^{b_{\scalebox{0.45}{$R$}}C}=p_i^{-1}{\rm Tr}_{Ab_{\scalebox{0.45}{$L$}}}\left[\langle i|^{b_{\scalebox{0.45}{$0$}}}\Gamma\rho\Gamma^\dagger|i\rangle^{b_{\scalebox{0.45}{$0$}}}\right].\nonumber
\end{eqnarray}
Condition (\ref{eq:gamma2dagger}) implies that
\begin{align}
{\rm supp}\left(\sum_ip_i\proj{i}^{b_{\scalebox{0.45}{$0$}}}\otimes\rho_i^{Ab_{\scalebox{0.45}{$L$}}}\otimes\rho_i^{b_{\scalebox{0.45}{$R$}}C}\right)\subseteq{\rm img}\:\Gamma.\nonumber
\end{align}
Therefore, due to Theorem \ref{thm:markovdechayden}, $T_{\Gamma,\{|i\rangle\}}({\rho}^{ABC})$ is a normalized Markov state conditioned by $B$ for any $\rho^{ABC}$.  

Suppose there exist three Hilbert spaces ${\mathcal H}^{b_{\scalebox{0.45}{$0$}}}$, ${\mathcal H}^{b_{\scalebox{0.45}{$L$}}}$, ${\mathcal H}^{b_{\scalebox{0.45}{$R$}}}$ and a linear isometry $\Gamma:{\mathcal H}^B\rightarrow{\mathcal H}^{b_{\scalebox{0.45}{$0$}}}\otimes{\mathcal H}^{b_{\scalebox{0.45}{$L$}}}\otimes{\mathcal H}^{b_{\scalebox{0.45}{$R$}}}$ that satisfy (\ref{eq:gamma2dagger}) and 
\begin{eqnarray}
\left\|\rho^{ABC}-T_{\Gamma,\{|i\rangle\}}({\rho}^{ABC})\right\|_1\leq\epsilon.\nonumber
\end{eqnarray}
It immediately follows that there exists a Markov state $\Upsilon^{ABC}(=T_{\Gamma,\{|i\rangle\}}({\rho}^{ABC}))$ that satisfy Inequality (\ref{eq:appmarkov}).

Conversely, suppose there exists a Markov state $\Upsilon^{ABC}$ that satisfy (\ref{eq:appmarkov}). Let $\Gamma:{\mathcal H}_\Upsilon^B\rightarrow{\mathcal H}^{b_{\scalebox{0.45}{$0$}}}\otimes{\mathcal H}^{b_{\scalebox{0.45}{$L$}}}\otimes{\mathcal H}^{b_{\scalebox{0.45}{$R$}}}$ be a Markov isometry on $B$ with respect to $\Upsilon^{ABC}$, and let
\begin{eqnarray}
\Upsilon_{M\!k}:=\Gamma^B\Upsilon^{ABC}\Gamma^{\dagger B}=\sum_iq_i\proj{i}^{b_{\scalebox{0.45}{$0$}}}\otimes\sigma_i^{Ab_{\scalebox{0.45}{$L$}}}\otimes\phi_i^{b_{\scalebox{0.45}{$R$}}C}\label{eq:upsilonmarkdec}
\end{eqnarray}
be a Markov decomposition of $\Upsilon^{ABC}$. Due to Equality (7) in \cite{waka15_markov_paper}, $\Gamma$ satisfies (\ref{eq:gamma2dagger}). We first assume ${\mathcal H}_\Upsilon^B={\mathcal H}^B$ for simplicity, and prove
\begin{eqnarray}
\left\|\rho^{ABC}-T_{\Gamma,\{|i\rangle\}}({\rho}^{ABC})\right\|_1\leq6\epsilon.\label{eq:decifmark}
\end{eqnarray}
By the triangle inequality, we have
\begin{eqnarray}
&&\left\|\rho^{ABC}-T_{\Gamma,\{|i\rangle\}}({\rho}^{ABC})\right\|_1\nonumber\\
&\leq&\left\|\rho^{ABC}-\Upsilon^{ABC}\right\|_1+\left\|\Upsilon^{ABC}-T_{\Gamma,\{|i\rangle\}}({\rho}^{ABC})\right\|_1\nonumber\\
&\leq&\left\|T_{\Gamma,\{|i\rangle\}}({\rho}^{ABC})-\Upsilon^{ABC}\right\|_1+\epsilon.\label{eq:first}
\end{eqnarray}
Next, (\ref{eq:markovdecrho}) and (\ref{eq:upsilonmarkdec}) imply that
\begin{eqnarray}
&&\left\|T_{\Gamma,\{|i\rangle\}}({\rho}^{ABC})-\Upsilon^{ABC}\right\|_1\nonumber\\
&=&\left\|\Gamma\left(T_{\Gamma,\{|i\rangle\}}({\rho}^{ABC})\right)\Gamma^\dagger-\Upsilon_{M\!k}\right\|_1\nonumber\\
&=&\sum_i\left\|p_i\rho_i^{Ab_{\scalebox{0.45}{$L$}}}\otimes\rho_i^{b_{\scalebox{0.45}{$R$}}C}-q_i\sigma_i^{Ab_{\scalebox{0.45}{$L$}}}\otimes\phi_i^{b_{\scalebox{0.45}{$R$}}C}\right\|_1\nonumber\\
&\leq&\sum_i\left\|p_i\rho_i^{Ab_{\scalebox{0.45}{$L$}}}\otimes\rho_i^{b_{\scalebox{0.45}{$R$}}C}-p_i\sigma_i^{Ab_{\scalebox{0.45}{$L$}}}\otimes\phi_i^{b_{\scalebox{0.45}{$R$}}C}\right\|_1\nonumber\\
&&+\sum_i\left\|p_i\sigma_i^{Ab_{\scalebox{0.45}{$L$}}}\otimes\phi_i^{b_{\scalebox{0.45}{$R$}}C}-q_i\sigma_i^{Ab_{\scalebox{0.45}{$L$}}}\otimes\phi_i^{b_{\scalebox{0.45}{$R$}}C}\right\|_1\nonumber\\
&=&\sum_ip_i\left\|\rho_i^{Ab_{\scalebox{0.45}{$L$}}}\otimes\rho_i^{b_{\scalebox{0.45}{$R$}}C}-\sigma_i^{Ab_{\scalebox{0.45}{$L$}}}\otimes\phi_i^{b_{\scalebox{0.45}{$R$}}C}\right\|_1\nonumber\\
&&+\sum_i|p_i-q_i|.\label{eq:second}
\end{eqnarray}
In addition, we have
\begin{eqnarray}
&&p_i\left\|\rho_i^{Ab_{\scalebox{0.45}{$L$}}}\otimes\rho_i^{b_{\scalebox{0.45}{$R$}}C}-\sigma_i^{Ab_{\scalebox{0.45}{$L$}}}\otimes\phi_i^{b_{\scalebox{0.45}{$R$}}C}\right\|_1\nonumber\\
&\leq&p_i\left\|\rho_i^{Ab_{\scalebox{0.45}{$L$}}}\otimes\rho_i^{b_{\scalebox{0.45}{$R$}}C}-\sigma_i^{Ab_{\scalebox{0.45}{$L$}}}\otimes\rho_i^{b_{\scalebox{0.45}{$R$}}C}\right\|_1\nonumber\\
&&+p_i\left\|\sigma_i^{Ab_{\scalebox{0.45}{$L$}}}\otimes\rho_i^{b_{\scalebox{0.45}{$R$}}C}-\sigma_i^{Ab_{\scalebox{0.45}{$L$}}}\otimes\phi_i^{b_{\scalebox{0.45}{$R$}}C}\right\|_1\nonumber\\
&=&p_i\left\|\rho_i^{Ab_{\scalebox{0.45}{$L$}}}-\sigma_i^{Ab_{\scalebox{0.45}{$L$}}}\right\|_1+p_i\left\|\rho_i^{b_{\scalebox{0.45}{$R$}}C}-\phi_i^{b_{\scalebox{0.45}{$R$}}C}\right\|_1\nonumber\\
&\leq&2p_i\left\|\rho_i^{Ab_{\scalebox{0.45}{$L$}}b_{\scalebox{0.45}{$R$}}C}-\sigma_i^{Ab_{\scalebox{0.45}{$L$}}}\otimes\phi_i^{b_{\scalebox{0.45}{$R$}}C}\right\|_1\nonumber\\
&\leq&2\left\|p_i\rho_i^{Ab_{\scalebox{0.45}{$L$}}b_{\scalebox{0.45}{$R$}}C}-q_i\sigma_i^{Ab_{\scalebox{0.45}{$L$}}}\otimes\phi_i^{b_{\scalebox{0.45}{$R$}}C}\right\|_1\nonumber\\
&&+2\left\|q_i\sigma_i^{Ab_{\scalebox{0.45}{$L$}}}\otimes\phi_i^{b_{\scalebox{0.45}{$R$}}C}-p_i\sigma_i^{Ab_{\scalebox{0.45}{$L$}}}\otimes\phi_i^{b_{\scalebox{0.45}{$R$}}C}\right\|_1\nonumber\\
&=&2\left\|p_i\rho_i^{Ab_{\scalebox{0.45}{$L$}}b_{\scalebox{0.45}{$R$}}C}-q_i\sigma_i^{Ab_{\scalebox{0.45}{$L$}}}\otimes\phi_i^{b_{\scalebox{0.45}{$R$}}C}\right\|_1+2|p_i-q_i|,\label{eq:third}
\end{eqnarray}
where we define
\begin{eqnarray}
\rho_i^{Ab_{\scalebox{0.45}{$L$}}b_{\scalebox{0.45}{$R$}}C}=p_i^{-1}\langle i|^{b_{\scalebox{0.45}{$0$}}}\Gamma\rho\Gamma^\dagger|i\rangle^{b_{\scalebox{0.45}{$0$}}}.\nonumber
\end{eqnarray}
Consider a state $\hat\rho$ defined by 
\begin{eqnarray}
{\hat\rho}:=\sum_ip_i\proj{i}^{b_{\scalebox{0.45}{$0$}}}\otimes\rho_i^{Ab_{\scalebox{0.45}{$L$}}b_{\scalebox{0.45}{$R$}}C},\nonumber
\end{eqnarray}
and let ${\mathcal D}^{b_{\scalebox{0.45}{$0$}}}$ be the completely dephasing operation on $b_0$ with respect to the basis $|i\rangle$.
We have ${\hat\rho}={\mathcal D}^{b_{\scalebox{0.45}{$0$}}}(\Gamma\rho\Gamma^\dagger)$, as well as $\Upsilon_{M\!k}={\mathcal D}^{b_{\scalebox{0.45}{$0$}}}(\Upsilon_{M\!k})$ from (\ref{eq:upsilonmarkdec}). Therefore, by the monotonicity of the trace distance,
\begin{eqnarray}
\left\|{\hat\rho}-\Upsilon_{M\!k}\right\|_1\leq\epsilon\label{eq:rhohatmark}
\end{eqnarray}
holds from (\ref{eq:appmarkov}), which leads to
\begin{eqnarray}
\sum_i\left\|p_i\rho_i^{Ab_{\scalebox{0.45}{$L$}}b_{\scalebox{0.45}{$R$}}C}-q_i\sigma_i^{Ab_{\scalebox{0.45}{$L$}}}\otimes\phi_i^{b_{\scalebox{0.45}{$R$}}C}\right\|_1\leq\epsilon.\label{eq:forth}
\end{eqnarray}
By tracing out $Ab_Lb_RC$ in (\ref{eq:rhohatmark}), we obtain
\begin{eqnarray}
\sum_i|p_i-q_i|\leq\epsilon.\label{eq:fifth}
\end{eqnarray}
Combining (\ref{eq:first}), (\ref{eq:second}), (\ref{eq:third}), (\ref{eq:forth}) and (\ref{eq:fifth}), we obtain (\ref{eq:decifmark}). 

An inequality similar to (\ref{eq:decifmark}) is obtained when ${\mathcal H}_\Upsilon^B\neq{\mathcal H}^B$ as well. Note that $\rho^{ABC}$ is invariant when projected onto the support of $\Upsilon^B$ up to a small error $2\sqrt{\epsilon}$, due to Inequality (\ref{eq:appmarkov}) and the gentle measurement lemma (see Lemma 9.4.1 in \cite{wildetext}).\\
\hfill$\blacksquare$

\section{M-Dec Cost of Pure States}\label{app:marksingle}

In this section, we summarize a result obtained in \cite{waka15_markov_paper} regarding a single-letter formula for the M-Dec cost of pure states. Let us first present a decomposition of a Hilbert space called the {\it Koashi-Imoto (KI)} decomposition, which is first introduced in \cite{koashi02} and is extended in \cite{hayden04}.

\begin{thm}\label{thm:kidec}(\!\!\cite{koashi02,hayden04}, see also Definition 3 and Lemma 4 in \cite{waka15_markov_paper})
Consider a quantum system $A$ and $A'$ described by a finite dimensional Hilbert space ${\mathcal H}^A$ and ${\mathcal H}^{A'}$, respectively. Associated to any bipartite quantum state $\Psi^{AA'}\in{\mathcal S}({\mathcal H}^A\otimes{\mathcal H}^{A'})$, there exist three Hilbert spaces ${\mathcal H}^{a_{\scalebox{0.45}{$0$}}}$, ${\mathcal H}^{a_{\scalebox{0.45}{$L$}}}$, ${\mathcal H}^{a_{\scalebox{0.45}{$R$}}}$ and an isometry $\Gamma$ from ${\mathcal H}_\Psi^A:={\rm supp}[\Psi^A]$ to ${\mathcal H}^{a_{\scalebox{0.45}{$0$}}}\otimes{\mathcal H}^{a_{\scalebox{0.45}{$L$}}}\otimes{\mathcal H}^{a_{\scalebox{0.45}{$R$}}}$ such that the following two properties hold.

\begin{enumerate}
\item
$\Gamma$ gives
\begin{eqnarray}
\Psi^{AA'}_{K\!I}:=\Gamma^A\Psi^{AA'}\Gamma^{\dagger A}=\sum_{j\in J}p_j\proj{j}^{a_{\scalebox{0.45}{$0$}}}\otimes\omega_j^{a_{\scalebox{0.45}{$L$}}}\otimes\varphi_j^{a_{\scalebox{0.45}{$R$}}A'}\!\!\!\!\!\!\!\!\!\!\nonumber\\
\label{eq:koashi02ofbistate}
\end{eqnarray}
with some probability distribution $\{p_{j}\}_{j\in J}$, orthonormal basis $\{\ket{j}\}_{j\in J}$ of ${\mathcal H}^{a_{\scalebox{0.45}{$0$}}}$, states $\omega_j\in{\mathcal S}({\mathcal H}^{a_{\scalebox{0.45}{$L$}}})$ and $\varphi_j\in{\mathcal S}({\mathcal H}^{a_{\scalebox{0.45}{$R$}}}\otimes{\mathcal H}^{A'})$.
\item A quantum operation $\mathcal E$ on ${\mathcal S}({\mathcal H}^A_\Psi)$ leaves $\Psi^{AA'}$ invariant if and only if there exists an isometry $U:{\mathcal H}^A_{\Psi}\rightarrow{\mathcal H}^A_{\Psi}\otimes{\mathcal H}^E$ such that a Stinespring dilation of $\mathcal E$ is given by ${\mathcal E}(\tau)={\rm Tr}_E[U\tau U^{\dagger}]$, and that $U$ is decomposed by $\Gamma$ as
\begin{eqnarray}
(\Gamma\otimes I^E)U\Gamma^\dagger=\sum_{j\in J}\proj{j}^{a_{\scalebox{0.45}{$0$}}}\otimes U_j^{a_{\scalebox{0.45}{$L$}}}\otimes I_j^{a_{\scalebox{0.45}{$R$}}}.\nonumber
\end{eqnarray}
Here, $I_j$ are the identity operator on ${\mathcal H}_j^{a_{\scalebox{0.45}{$R$}}}:={\rm supp}\sum_k\rho_{j|k}$, and $U_j:{\mathcal H}_j^{a_{\scalebox{0.45}{$L$}}}\rightarrow{\mathcal H}_j^{a_{\scalebox{0.45}{$L$}}}\otimes{\mathcal H}^{E}$ are isometries that satisfy ${\rm Tr}_E[U_j\omega_jU_j^{\dagger}]=\omega_j$ for all $j$, where ${\mathcal H}_j^{a_{\scalebox{0.45}{$L$}}}:={\rm supp}\:\omega_{j}$.
\end{enumerate}
\end{thm}
We call $\Gamma$ as the {\it KI isometry on system $A$ with respect to $\Psi^{AA'}$}, and (\ref{eq:koashi02ofbistate}) as the {\it KI decomposition of $\Psi^{AA'}$ on $A$}. The KI isometry and the KI decomposition are uniquely determined from $\Psi^{AA'}$, up to trivial changes of the basis. 

A single-letter formula for the M-Dec cost of tripartite pure states is obtained based on the KI decomposition.

\begin{thm}\label{thm:strongmarkcostequality}(Theorem 8 in \cite{waka15_markov_paper})
Let $|\Psi\rangle^{ABC}$ be a pure state, and let
\begin{eqnarray}
\Psi_{K\!I}^{AC}=\sum_{j\in J}p_j\proj{j}^{a_{\scalebox{0.45}{$0$}}}\otimes\omega_j^{a_{\scalebox{0.45}{$L$}}}\otimes\varphi_j^{a_{\scalebox{0.45}{$R$}}C}\nonumber
\label{eq:kidecofpsi}
\end{eqnarray}
be the KI decomposition of $\Psi^{AC}$ on $A$. Then we have
\begin{eqnarray}
M_{A|B}^D(\Psi^{ABC})=H(\{p_j\}_{j\in J})+2\sum_{j\in J}p_jS(\varphi_j^{a_{\scalebox{0.45}{$R$}}}).\nonumber
\end{eqnarray}
\end{thm}
As we proved in \cite{waka15_markov_paper} (see Appendix B-B therein), the error $\epsilon$ vanishes exponentially with $n$. Thus Theorem \ref{thm:strongmarkcostequality} holds even when we additionally require in Definition \ref{dfn:mofpsi} that $\lim_{\epsilon\rightarrow0}n\cdot\epsilon_n=0$.

\section{Proof of Theorem \ref{thm:leftrightmark}}\label{prf:leftrightmark}

In this Appendix, we present a proof of Theorem \ref{thm:leftrightmark}. Proofs of Inequalities (\ref{eq:leftmark}) and (\ref{eq:rightmark}) proceeds almost in parallel.  Let us start with a summary of the continuity bounds of quantum entropies and mutual informations.

\subsection{Continuity of Quantum Entropies}\label{app:content}

Define
\begin{eqnarray}
\eta_0(x):=\begin{cases}
-x\log{x}&(x\leq 1/e)\\
\frac{1}{e}&(x\geq 1/e)
\end{cases},\nonumber
\end{eqnarray}
$\eta(x)=x+\eta_0(x)$ and $h(x):=\eta_0(x)+\eta_0(1-x)$, where $e$ is the base of natural logarithm. For two states $\rho$ and $\sigma$ in a $d$-dimensional quantum system ($d<\infty$) such that $\|\rho-\sigma\|_1\leq\epsilon$, we have
\begin{eqnarray}
|S(\rho)-S(\sigma)|\leq\epsilon\log{d}+\eta_0(\epsilon)\leq\eta(\epsilon)\log{d},\label{eq:fannes}
\end{eqnarray}
which is called the {\it Fannes inequality}\cite{fannes73}. For two bipartite states $\rho,\sigma\in{\mathcal S}({\mathcal H}^A\otimes{\mathcal H}^B)$ such that  $\|\rho-\sigma\|_1\leq\epsilon<1$, we have
\begin{eqnarray}
|S(A|B)_\rho-S(A|B)_\sigma|&\leq&4\epsilon\log{d_A}+2h(\epsilon)\nonumber\\
&\leq&4\eta(\epsilon)\log{d_A},\label{eq:alicki}
\end{eqnarray}
which is called the {\it Alicki-Fannes inequality}\cite{alicki04}. Note that the upper bound in (\ref{eq:alicki}) does not depend on $d_B$. As a consequence, we have
\begin{eqnarray}
|I(A:B)_\rho-I(A:B)_\sigma|\leq5\eta(\epsilon)\log{d_A}.\label{eq:contmii}
\end{eqnarray}

\subsection{Proof of Inequality (\ref{eq:leftmark})}\label{sec:prfleftmark}

We prove the first inequality in (\ref{eq:leftmark}) by showing that any $R$ satisfying $R>M^R_{A|BC}(\rho^{ABC})$ also satisfies $R\geq I(A:C|B)$. By definition, for an arbitrary $R>M^R_{A|BC}(\rho^{ABC})$, $\epsilon\in(0,1)$ and sufficiently large $n$, there exists a random unitary operation ${\mathcal V}_n:\tau\mapsto2^{-nR}\sum_{k=1}^{2^{nR}}V_{n,k}\tau V_{n,k}^{\dagger}$ on ${\bar A}$ and a quantum operation ${\mathcal R}_n:{\bar B}\rightarrow {\bar A}{\bar B}$ that satisfy
\begin{eqnarray}
\left\|{\mathcal V}_n((\rho^{ABC})^{\otimes n})-{\mathcal R}_n((\rho^{BC})^{\otimes n})\right\|_1\leq\epsilon.\label{eq:defmarkovianizing22}
\end{eqnarray}
Let $|\psi\rangle^{ABCD}$ be a purification of $\rho^{ABC}$, $E$ be a quantum system with dimension $2^{nR}$, and let $\{|k\rangle\}_{k=1}^{2^{nR}}$ be an orthonormal basis of ${\mathcal H}^E$. Defining an isometry $W:{\bar A}\rightarrow E{\bar A}$ by $W=\sum_{k=1}^{2^{nR}}|k\rangle^{E}\otimes V^{{\bar A}}_{n,k}$, a Stinespring dilation of ${\mathcal V}_n$ is given by ${\mathcal V}_n(\tau)={\rm Tr}_E[W\tau W^\dagger]$. Then a purification of $\rho_n'^{ABC}:={\mathcal V}_n((\rho^{ABC})^{\otimes n})$ is given by $|\psi_n'\rangle^{E{\bar A}{\bar B}{\bar C}{\bar D}}:=W(|\psi\rangle^{ABCD})^{\otimes n}$. For this state, we have
\begin{eqnarray}
nR&\geq&S(E)_{\psi'_n}\nonumber\\
&=&S({\bar A}{\bar B}{\bar C}{\bar D})_{\psi_n'}\nonumber\\
&\geq&S({\bar A}{\bar B}{\bar C})_{\psi_n'}-S({\bar D})_{\psi_n'}\nonumber\\
&=&S({\bar A}{\bar B}{\bar C})_{\rho_n'}-S({\bar D})_{\psi^{\otimes n}}\nonumber\\
&=&S({\bar A}{\bar B}{\bar C})_{\rho_n'}-nS(ABC)_{\rho},\label{eq:qcmient01}
\end{eqnarray}
where the third line follows by the Araki-Lieb inequality\cite{araki}. The first term satisfies
\begin{eqnarray}
&&S({\bar A}{\bar B}{\bar C})_{\rho_n'}\nonumber\\
&=&S({\bar C}|{\bar A}{\bar B})_{\rho_n'}+S({\bar A}{\bar B})_{\rho_n'}\nonumber\\
&\geq&S({\bar C}|{\bar A}{\bar B})_{\rho_n'}+S({\bar A}{\bar B})_{\rho^{\otimes n}}\nonumber\\
&=&S({\bar C}|{\bar A}{\bar B})_{\rho_n'}+nS(AB)_{\rho},\;\;\;\;\label{eq:qcmient11}
\end{eqnarray}
where the third line follows because the von Neumann entropy is nondecreasing under random unitary operations. Define $\rho_n''^{ABC}:={\mathcal R}_n((\rho^{BC})^{\otimes n})$. Note also that
\begin{eqnarray}
&&S({\bar C}|{\bar A}{\bar B})_{\rho_n'}\nonumber\\
&\geq&S({\bar C}|{\bar A}{\bar B})_{\rho_n''}-4n\eta(\epsilon)\log{d_C}\nonumber\\
&\geq&S({\bar C}|{\bar B})_{\rho^{\otimes n}}-4n\eta(\epsilon)\log{d_C}\nonumber\\
&=&n\left(S(BC)_{\rho}-S(B)_{\rho}\right)-4n\eta(\epsilon)\log{d_C},\;\;\;\label{eq:qcmient21}
\end{eqnarray}
where the second line follows from (\ref{eq:defmarkovianizing22}) and (\ref{eq:alicki}), the third line by the data processing inequality, and the fourth line because
\begin{eqnarray}
\rho_n'^{{\bar B}{\bar C}}={\rm Tr}_{{\bar A}}[{\mathcal V}_n((\rho^{ABC})^{\otimes n})]=(\rho^{BC})^{\otimes n}.\nonumber
\end{eqnarray}
From (\ref{eq:qcmient01}), (\ref{eq:qcmient11}) and (\ref{eq:qcmient21}), we obtain
\begin{eqnarray}
R\geq I(A:C|B)_\rho-4\eta(\epsilon)\log{d_C}.\nonumber
\end{eqnarray}
Since this relation holds for any $R>M^R_{A|BC}(\rho^{ABC})$ and $\epsilon>0$, we have the first inequality in (\ref{eq:leftmark}).

The proof for the second inequality is as follows. For any $R>M_{A|B}^D(\rho^{ABC})$, $\epsilon>0$ and sufficiently large $n$, there exists a random unitary operation ${\mathcal V}_n:\tau\mapsto2^{-nR}\sum_{k=1}^{2^{nR}}V_{n,k}\tau V_{n,k}^{\dagger}$ on ${\bar A}$ and a Markov state $\Upsilon^{{\bar A}{\bar B}{\bar C}}$ conditioned by ${\bar B}$ that satisfy
\begin{eqnarray}
\left\|{\mathcal V}_n(\rho^{\otimes n})-\Upsilon^{{\bar A}{\bar B}{\bar C}}\right\|_1\leq\frac{\epsilon}{2}.\label{eq:markov2}
\end{eqnarray}
Let ${\mathcal R}_n:{\bar B}\rightarrow {\bar A}{\bar B}$ be a quantum operation that satisfy
\begin{eqnarray}
\Upsilon^{{\bar A}{\bar B}{\bar C}}={\mathcal R}_n(\Upsilon^{{\bar B}{\bar C}}).\nonumber
\end{eqnarray}
By tracing out ${\bar A}$ in (\ref{eq:markov2}), we have
\begin{eqnarray}
\left\|(\rho^{BC})^{\otimes n}-\Upsilon^{{\bar B}{\bar C}}\right\|_1\leq\frac{\epsilon}{2},\nonumber
\end{eqnarray}
and consequently,
\begin{eqnarray}
\left\|{\mathcal R}_n\left((\rho^{BC})^{\otimes n}\right)-\Upsilon^{{\bar A}{\bar B}{\bar C}}\right\|_1\leq\frac{\epsilon}{2}.\nonumber
\end{eqnarray}
Therefore, by the triangle inequality, we obtain
\begin{eqnarray}
\left\|{\mathcal V}_n(\rho^{\otimes n})-{\mathcal R}_n\left((\rho^{BC})^{\otimes n}\right)\right\|_1\leq\epsilon,\label{eq:matsuya}
\end{eqnarray}
which implies $R\geq M^R_{A|BC}(\rho^{ABC})$. Thus we have the second inequality in (\ref{eq:leftmark}).\hfill$\blacksquare$

\subsection{Proof of Inequality (\ref{eq:rightmark})}

For an arbitrary $R>M^R_{A|AB}(\rho^{ABC})$, $\epsilon\in(0,1)$ and sufficiently large $n$, there exists a random unitary operation ${\mathcal V}_n:\tau\mapsto2^{-nR}\sum_{k=1}^{2^{nR}}V_{n,k}\tau V_{n,k}^{\dagger}$ on ${\bar A}$ and a linear CPTP map ${\mathcal R}_n:{\bar B}\rightarrow {\bar B}{\bar C}$ that satisfy
\begin{eqnarray}
\left\|{\mathcal V}_n((\rho^{ABC})^{\otimes n})-({\mathcal V}_n\otimes{\mathcal R}_n)((\rho^{AB})^{\otimes n})\right\|_1\leq\epsilon.\label{eq:defmarkovianizing222}
\end{eqnarray}
Define states $|\psi\rangle^{ABCD}$, $\rho_n'^{ABC}$ and $|\psi_n'\rangle^{E{\bar A}{\bar B}{\bar C}{\bar D}}$ in the same way as in Appendix \ref{sec:prfleftmark}. For these states, in addition to (\ref{eq:qcmient01}), we have
\begin{eqnarray}
&&S({\bar A}{\bar B}{\bar C})_{\rho_n'}\nonumber\\
&=&S({\bar A}|{\bar B}{\bar C})_{\rho_n'}+S({\bar B}{\bar C})_{\rho_n'}\nonumber\\
&=&S({\bar A}|{\bar B}{\bar C})_{\rho_n'}+nS(BC)_{\rho},\;\;\;\;\label{eq:qcmient1}
\end{eqnarray}
where the third line follows from $\rho_n'^{{\bar B}{\bar C}}=(\rho^{BC})^{\otimes n}$. Using (\ref{eq:defmarkovianizing222}), it holds that
\begin{eqnarray}
&&S({\bar A}|{\bar B}{\bar C})_{\rho_n'}\nonumber\\
&\geq&S({\bar A}|{\bar B}{\bar C})_{{\mathcal R}_n(\rho_n')}-4n\eta(\epsilon)\log{d_A}\nonumber\\
&\geq&S({\bar A}|{\bar B})_{\rho_n'}-4n\eta(\epsilon)\log{d_A}\nonumber\\
&=&S({\bar A}{\bar B})_{\rho_n'}-S({\bar B})_{\rho_n'}-4n\eta(\epsilon)\log{d_A}\nonumber\\
&\geq&S({\bar A}{\bar B})_{\rho^{\otimes n}}-S({\bar B})_{\rho^{\otimes n}}-4n\eta(\epsilon)\log{d_A}\nonumber\\
&=&n\left(S(AB)_{\rho}-S(B)_{\rho}\right)-4n\eta(\epsilon)\log{d_A}.\label{eq:qcmient2}
\end{eqnarray}
Here, the second line follows from (\ref{eq:defmarkovianizing222}) and (\ref{eq:alicki}); the third line by the data processing inequality; and the fifth line by the von Neumann entropy being nondecreasing under random unitary operations, in addition to $\rho_n'^{{\bar B}}=(\rho^{B})^{\otimes n}$. From (\ref{eq:qcmient01}), (\ref{eq:qcmient1}) and (\ref{eq:qcmient2}), we obtain
\begin{eqnarray}
R\geq I(A:C|B)_\rho-4\eta(\epsilon)\log{d_A},\nonumber
\end{eqnarray}
which concludes the proof for the first inequality in (\ref{eq:rightmark}).

The second inequality is proved as follows. Consider Inequality (\ref{eq:markov2}), and let ${\mathcal R}_n:{\bar B}\rightarrow {\bar B}{\bar C}$ be a linear CPTP map that satisfy
\begin{eqnarray}
\Upsilon^{{\bar A}{\bar B}{\bar C}}={\mathcal R}_n(\Upsilon^{{\bar A}{\bar B}}).\nonumber
\end{eqnarray}
By tracing out ${\bar C}$ in (\ref{eq:markov2}), we have
\begin{eqnarray}
\left\|{\mathcal V}_n\left((\rho^{AB})^{\otimes n}\right)-\Upsilon^{{\bar A}{\bar B}}\right\|_1\leq\frac{\epsilon}{2},\nonumber
\end{eqnarray}
which implies
\begin{eqnarray}
\left\|({\mathcal V}_n\otimes{\mathcal R}_n)\left((\rho^{AB})^{\otimes n}\right)-\Upsilon^{{\bar A}{\bar B}{\bar C}}\right\|_1\leq\frac{\epsilon}{2}.\nonumber
\end{eqnarray}
Therefore, by the triangle inequality, we obtain
\begin{eqnarray}
\left\|{\mathcal V}_n\left((\rho^{ABC})^{\otimes n}\right)-({\mathcal V}_n\otimes{\mathcal R}_n)\left((\rho^{AB})^{\otimes n}\right)\right\|_1\leq\epsilon,\label{eq:sukiya}
\end{eqnarray}
which implies $R\geq M^R_{A|AB}(\rho^{ABC})$. Thus we have the second inequality in (\ref{eq:rightmark}). 
\hfill$\blacksquare$

\section{Proof of Theorem \ref{thm:leftmarkpure}}\label{prf:leftmarkpure}

In this Appendix, we provide a rigorous proof of Theorem \ref{thm:leftmarkpure}. We first prove Lemma \ref{thm:corrtran}. We then prove Equalities (\ref{eq:leftmarkpure}) and (\ref{eq:rightmarkpure}) by using the obtained result. In the following, we denote systems ${\bar A}$, ${\bar B}$ and ${\bar C}$ by $\bar A$, $\bar B$ and $\bar C$ for simplicity of notation. We informally denote the composite systems $a_{\scalebox{0.6}{$0$}}a_{\scalebox{0.6}{$L$}}a_{\scalebox{0.6}{$R$}}$ by $A$ and $b_{\scalebox{0.6}{$0$}}b_{\scalebox{0.6}{$L$}}b_{\scalebox{0.6}{$R$}}$ by $B$, when there is no fear of confusion.

\subsection{Proof of Lemma \ref{thm:corrtran}}\label{app:prfcortran}

Let $\Gamma$ be the KI isometry on $A$ with respect to $\Psi^{AC}$, and suppose the KI decomposition of $\Psi^{AC}$ on $A$ is given by 
\begin{eqnarray}
\Psi^{AC}_{K\!I}:=\Gamma^A\Psi^{AC}\Gamma^{\dagger A}=\sum_{j\in J}p_j\proj{j}^{a_{\scalebox{0.45}{$0$}}}\otimes\omega_j^{a_{\scalebox{0.45}{$L$}}}\otimes\varphi_j^{a_{\scalebox{0.45}{$R$}}C}.\nonumber
\end{eqnarray}
As we prove in \cite{waka15_markov_paper} (see Lemma 10 therein), there exist three Hilbert spaces ${\mathcal H}^{b_{\scalebox{0.45}{$0$}}}$, ${\mathcal H}^{b_{\scalebox{0.45}{$L$}}}$, ${\mathcal H}^{b_{\scalebox{0.45}{$R$}}}$ and an isometry $\Gamma':{\mathcal H}_\Psi^B\rightarrow {\mathcal H}^{b_{\scalebox{0.45}{$0$}}}\otimes{\mathcal H}^{b_{\scalebox{0.45}{$L$}}}\otimes{\mathcal H}^{b_{\scalebox{0.45}{$R$}}}$ such that $|\Psi\rangle^{ABC}$ is decomposed as
\begin{eqnarray}
|\Psi_{K\!I}\rangle&:=&(\Gamma^A\otimes\Gamma'^B)|\Psi\rangle^{ABC}\nonumber\\
&=&\sum_{j\in J}\sqrt{p_j}\ket{j}^{a_{\scalebox{0.45}{$0$}}}\ket{j}^{b_{\scalebox{0.45}{$0$}}}\ket{\omega_j}^{a_{\scalebox{0.45}{$L$}}b_{\scalebox{0.45}{$L$}}}\ket{\varphi_j}^{a_{\scalebox{0.45}{$R$}}b_{\scalebox{0.45}{$R$}}C},
\label{eq:kidecoftristate}
\end{eqnarray}
where $\ket{\omega_j}^{a_{\scalebox{0.45}{$L$}}b_L}$ and $\ket{\varphi_j}^{a_{\scalebox{0.45}{$R$}}b_{\scalebox{0.45}{$R$}}C}$ are purifications of $\omega_j^{a_{\scalebox{0.45}{$L$}}}$ and $\varphi_j^{a_{\scalebox{0.45}{$R$}}C}$, respectively, and $\inpro{j}{j'}^{b_{\scalebox{0.45}{$0$}}}=\delta_{jj'}$. Let $A_l$ denote the $l$-th copy of $A$ in ${\bar A}$. For ${\mathcal E}$ that satisfies (\ref{eq:cortranerror}), define a quantum channel on $A_l$ ($1\leq l\leq n$) by
\begin{eqnarray}
&&{\mathcal E}_l(\tau^{A_l})={\rm Tr}_{{\bar A}\setminus A_l}\left[{\mathcal E}\left(\Psi^{A_1}\otimes\cdots\otimes\Psi^{A_{l-1}}\otimes\tau^{A_l}\right.\right.\;\;\;\;\nonumber\\
&&\;\;\;\;\;\;\;\;\;\;\;\;\;\;\;\;\;\;\;\;\;\;\;\;\;\;\;\;\;\;\;\;\left.\left.\otimes\Psi^{A_{l+1}}\otimes\cdots\otimes\Psi^{A_{n}}\right)\right],\nonumber
\end{eqnarray}
where ${\rm Tr}_{{\bar A}\setminus A_l}$ denotes the partial trace over $A_1\cdots A_{l-1}A_{l+1}\cdots A_n$. From (\ref{eq:cortranerror}), we have
\begin{eqnarray}
\left\|{\mathcal E}_{l}(\Psi^{{A_l}{C_l}})-\Psi^{{A_l}{C_l}}\right\|_1\leq\epsilon\label{eq:cheesepizza}
\end{eqnarray}
for any $1\leq l\leq n$.

Define a quantum operation ${\mathcal F}$ on ${\mathcal S}({\mathcal H}_\Psi^B)$ and a state ${\tilde\Psi}^{ABC}$ by
\begin{eqnarray}
{\mathcal F}(\tau)=\Gamma'^{\dagger}\!\left(\sum_j\proj{j}^{{b}_{\scalebox{0.45}{$0$}}}{\rm Tr}_{{b}_{\scalebox{0.45}{$L$}}}[\Gamma'\tau\Gamma'^{\dagger}]\proj{j}^{{b}_{\scalebox{0.45}{$0$}}}\otimes\omega_j^{{b}_{\scalebox{0.45}{$L$}}}\!\right)\!\Gamma',\!\!\!\!\!\!\!\!\!\nonumber
\end{eqnarray} 
and ${\tilde\Psi}^{ABC}:={\mathcal F}^B(|\Psi\rangle\!\langle\Psi|)$. It immediately follows from (\ref{eq:kidecoftristate}) that
\begin{eqnarray}
&&\!\!\!\!\!\!\!\!{\tilde\Psi}_{K\!I}:=(\Gamma^A\otimes\Gamma'^B){\tilde\Psi}^{ABC}(\Gamma^A\otimes\Gamma'^B)^\dagger\nonumber\\
&&\!\!\!\!=\sum_{j\in J}p_j\proj{j}^{a_{\scalebox{0.45}{$0$}}}\otimes\proj{j}^{b_{\scalebox{0.45}{$0$}}}\otimes{\omega_j}^{a_{\scalebox{0.45}{$L$}}}\otimes\proj{\varphi_j}^{a_{\scalebox{0.45}{$R$}}b_{\scalebox{0.45}{$R$}}C}\otimes{\omega_j}^{b_{\scalebox{0.45}{$L$}}}.\nonumber\\\label{eq:kipure}
\end{eqnarray}
Define a function $\zeta_{{}_{\Psi}}\!(\epsilon)$ by
\begin{eqnarray}
&&\!\!\!\!\!\!\zeta_{{}_{\Psi}}\!(\epsilon):=\nonumber\\
&&\!\!\!\sup\left\{\left.\|{\mathcal G}({\tilde\Psi}^{ABC})-{\tilde\Psi}^{ABC}\|_1\right|\|{\mathcal G}(\Psi^{AC})-\Psi^{AC}\|_1\leq\epsilon\right\},\nonumber
\end{eqnarray}
where the supremum is taken over quantum operations $\mathcal G$ on $A$. As we proved in \cite{waka15_markov_paper} (see Appendix B-E therein), this function satisfies $\lim_{\epsilon\rightarrow0}\zeta_{{}_{\Psi}}\!(\epsilon)=0$. 

From (\ref{eq:cheesepizza}), we have
\begin{eqnarray}
\left\|{\mathcal E}_{l}({\tilde\Psi}^{A_lB_lC_l})-{\tilde\Psi}^{A_lB_lC_l}\right\|_1\leq\zeta_{{}_{\Psi}}\!(\epsilon)\nonumber
\end{eqnarray}
for any $1\leq l\leq n$. By Inequality (\ref{eq:contmii}), it follows that
\begin{eqnarray}
I(A:BC)_{\tilde\Psi}-I(A_l:B_lC_l)_{{\mathcal E}_l(\tilde\Psi)}\leq 5\eta(\zeta_{{}_{\Psi}}\!(\epsilon))\log{d_A},\nonumber
\end{eqnarray}
and consequently, that
\begin{eqnarray}
&&nI(A:BC)_{\tilde\Psi}-\sum_{l=1}^nI(A_l:B_lC_l)_{{\mathcal E}_l(\tilde\Psi)}\nonumber\\
&&\leq 5n\eta(\zeta_{{}_{\Psi}}\!(\epsilon))\log{d_A}.\label{eq:condentdif}
\end{eqnarray}
We also have
\begin{eqnarray}
&&I({\bar A}:{\bar B}{\bar C})_{{\mathcal E}({\tilde\Psi}^{\otimes n})}\nonumber\\
&=&S({\bar B}{\bar C})_{{\mathcal E}({\tilde\Psi}^{\otimes n})}-S({\bar B}{\bar C}|{\bar A})_{{\mathcal E}({\tilde\Psi}^{\otimes n})}\nonumber\\
&=&S({\bar B}{\bar C})_{{\tilde\Psi}^{\otimes n}}\nonumber\\
&&\;\;\;\;\;\;\;-\sum_{l=1}^nS(B_lC_l|{{\bar A}\:B_1C_1\cdots B_{l-1}C_{l-1}})_{{\mathcal E}({\tilde\Psi}^{\otimes n})}\nonumber\\
&\geq& \sum_{l=1}^nS(B_lC_l)_{{\tilde\Psi}}-\sum_{l=1}^nS(B_lC_l|A_l)_{{\mathcal E}({\tilde\Psi}^{\otimes n})}\nonumber\\
&=&\sum_{l=1}^nS(B_lC_l)_{{\mathcal E}_l({\tilde\Psi})}-\sum_{l=1}^nS(B_lC_l|A_l)_{{\mathcal E}_l({\tilde\Psi})}\nonumber\\
&=&\sum_{l=1}^nI(A_l:B_lC_l)_{{\mathcal E}_l({\tilde\Psi})}.
\label{eq:condentdif2}
\end{eqnarray}
Here, we used the fact that ${\mathcal E}$ on $\bar A$ does not change the reduced state on ${\bar B}{\bar C}$, and that 
\begin{eqnarray}
{\rm Tr}_{{\bar A}\setminus A_l,{\bar B}\setminus B_l,{\bar C}\setminus C_l}\left[{\mathcal E}\left({\tilde\Psi}^{\otimes{n}}\right)\right]={\mathcal E}_l({\tilde\Psi}^{A_lB_lC_l}),\nonumber
\end{eqnarray}
because of ${\tilde\Psi}^A_{l'}=\Psi^A_{l'}$. Combining (\ref{eq:condentdif}) and (\ref{eq:condentdif2}), we obtain
\begin{eqnarray}
nI(A:BC)_{\tilde\Psi}&\leq& I({\bar A}:{\bar B}{\bar C})_{{\mathcal E}({\tilde\Psi}^{\otimes n})}\nonumber\\
&&+5n\eta(\zeta_{{}_{\Psi}}\!(\epsilon))\log{d_A}.\nonumber
\end{eqnarray}

The L.H.S. in this inequality is computed from (\ref{eq:kipure}) and Theorem \ref{thm:strongmarkcostequality} as
\begin{eqnarray}
I(A:BC)_{\tilde\Psi}&=&H(\{p_j\}_{j\in J})+2\sum_{j\in J}p_jS(\varphi_j^{a_{\scalebox{0.45}{$R$}}})\nonumber\\
&=&M_{A|B}^D(\Psi^{ABC}).\nonumber
\end{eqnarray}
The data processing inequality yields
\begin{eqnarray}
I({\bar A}:{\bar B}{\bar C})_{{\mathcal E}({\tilde\Psi}^{\otimes n})}\leq I({\bar A}:{\bar B}{\bar C})_{{\mathcal E}({\Psi}^{\otimes n})}\nonumber
\end{eqnarray}
for the R.H.S. in (\ref{eq:kipure}). Thus we obtain
\begin{eqnarray}
&&\frac{1}{n}I({\bar A}:{\bar B}{\bar C})_{{\mathcal E}({\Psi}^{\otimes n})}\nonumber\\
&&\geq M_{A|B}^D(\Psi^{ABC})-5\eta(\zeta_{{}_{\Psi}}\!(\epsilon))\log{d_A},\nonumber
\end{eqnarray}
which completes the proof of Lemma 12.
\hfill$\blacksquare$

\subsection{Proof of Equality (\ref{eq:leftmarkpure})}
We prove $M^R_{A|BC}(\Psi^{ABC})\geq M^D_{A|B}(\Psi^{ABC})$, which, together with Inequality (\ref{eq:leftmark}), implies Equality (\ref{eq:leftmarkpure}). The proof presented here also provides an alternative proof for the converse part of Theorem 8 in \cite{waka15_markov_paper}.

For an arbitrary $R>M^R_{A|BC}(\Psi^{ABC})$, $\epsilon\in(0,1)$ and sufficiently large $n$, there exist a random unitary operation ${\mathcal V}_n:\tau\mapsto2^{-nR}\sum_{k=1}^{2^{nR}}V_{n,k}\tau V_{n,k}^{\dagger}$ on ${\bar A}$ and a quantum operation ${\mathcal R}_n:{\bar B}\rightarrow {\bar A}{\bar B}$ that satisfy
\begin{eqnarray}
\left\|{\mathcal V}_n((\Psi^{ABC})^{\otimes n})-{\mathcal R}_n((\Psi^{BC})^{\otimes n})\right\|_1\leq\epsilon.\label{eq:vnrnep}
\end{eqnarray}
Define an isometry $U_1:{\bar A}\rightarrow{\bar A}G$ by
\begin{eqnarray}
U_1:=\frac{1}{\sqrt{2^{nR}}}\sum_{k=1}^{2^{nR}}|k\rangle^G\otimes V^{\bar A}_{n,k},\nonumber
\end{eqnarray}
where $\{|k\rangle\}_{k=1}^{2^{nR}}$ is an orthonormal basis of ${\mathcal H}^G$. A purification of ${\mathcal V}_n((\Psi^{ABC})^{\otimes n})$ is then given by
\begin{eqnarray}
\!\!\!\!|\Psi_{{\mathcal V}_n}\rangle^{{\bar A}{\bar B}{\bar C}G}&:=&U_1|\Psi^{\otimes n}\rangle^{{\bar A}{\bar B}{\bar C}}\nonumber\\
&=&\frac{1}{\sqrt{2^{nR}}}\sum_{k=1}^{2^{nR}}|k\rangle^G\otimes V_{n,k}|\Psi^{\otimes n}\rangle^{{\bar A}{\bar B}{\bar C}}.\label{eq:kakugohaiika}
\end{eqnarray}
Let $E$ be an ancillary system with a sufficiently large dimension, $W:{\bar B}\rightarrow{\bar A}{\bar B}E$ be an isometry such that a Stinespring dilation of ${\mathcal R}_n$ is given by ${\mathcal R}_n(\tau)={\rm Tr}_E[W\tau W^\dagger]$, and let $A_c$ be a system which is identical to $A$. Then a purification of ${\mathcal R}_n((\Psi^{BC})^{\otimes n})$ is given by
\begin{eqnarray}
|\Psi_{{\mathcal R}_n}\rangle^{{\bar A}{\bar B}{\bar C}{\bar A}_cE}:=W|\Psi^{\otimes n}\rangle^{{\bar A}_c{\bar B}{\bar C}}.\label{eq:orehadekiteru}
\end{eqnarray} 
From (\ref{eq:vnrnep}), (\ref{eq:kakugohaiika}), (\ref{eq:orehadekiteru}) and Uhlmann's theorem\cite{uhlmann76}, there exists an isometry $U_2:G\rightarrow{\bar A}_cE$ such that
\begin{eqnarray}
&&\!\!\!\!\!\!\!\!\!\!\left\|U_2U_1|\Psi^{\otimes n}\rangle\!\langle\Psi^{\otimes n}|^{{\bar A}{\bar B}{\bar C}}U_1^\dagger U_2^\dagger-W|\Psi^{\otimes n}\rangle\!\langle\Psi^{\otimes n}|^{{\bar A}_c{\bar B}{\bar C}}W^\dagger|\right\|_1\nonumber\\
&&\!\!\!\!\!\!=\left\|U_2|\Psi_{{\mathcal V}_n}\rangle\!\langle\Psi_{{\mathcal V}_n}|U_2^\dagger-|\Psi_{{\mathcal R}_n}\rangle\!\langle\Psi_{{\mathcal R}_n}|\right\|_1\leq\epsilon.\label{eq:ariariariari}
\end{eqnarray}

\begin{figure}[t]
\begin{center}
\includegraphics[bb={0 0 356 522}, scale=0.4]{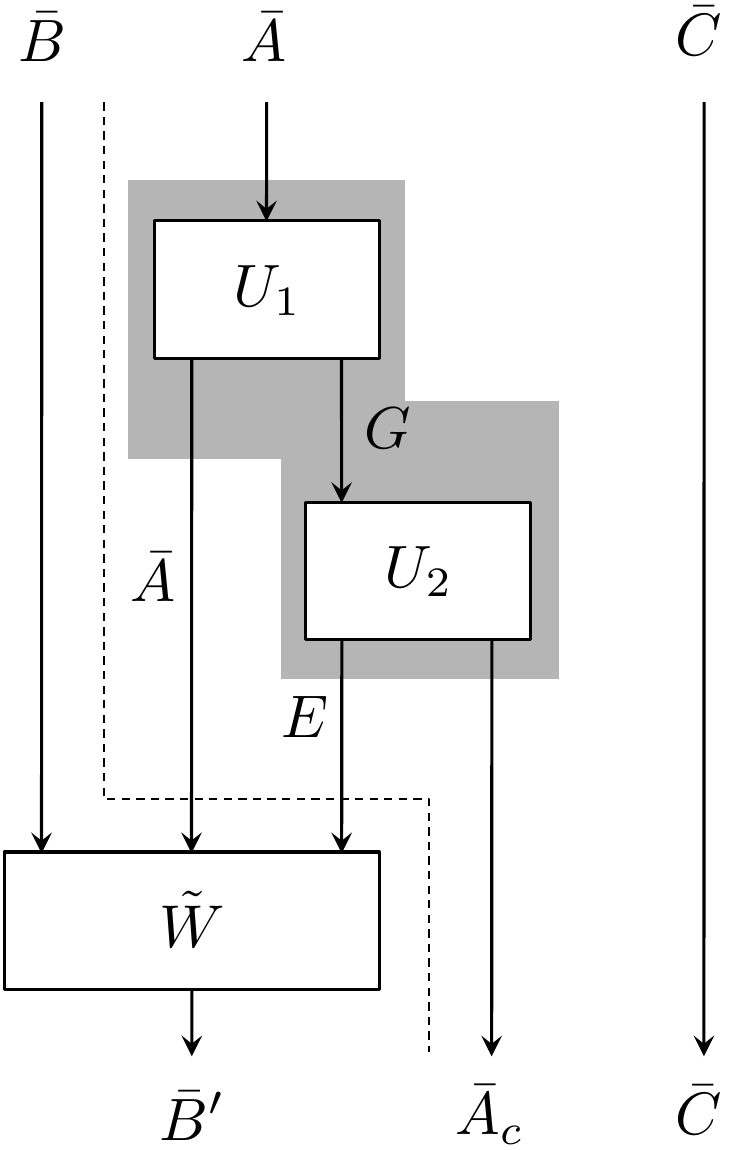}
\end{center}\caption{A graphical representation of the state transformation represented by a unitary isomorphism ${\tilde W}U_2U_1$ in Inequality (\ref{eq:uiuiiww}). ${\bar A}_c$ is identical to ${\bar A}$, and ${\bar B}'$ is a system represented by the ``extended'' Hilbert space ${\mathcal H}^{\bar B}\oplus{\mathcal H}_{\perp}$. Inequality (\ref{eq:uiuiiww}) states that $|\Psi^{\otimes n}\rangle$ is almost invariant by the action of this transformation. In particular, the reduced state on ${\bar A}_c{\bar C}$ of the final state is almost equal to that on ${\bar A}{\bar C}$ of the initial state. Discarding ${\bar B}'$ after applying $\tilde W$ is equivalent to discarding ${\bar B}$, ${\bar A}$ and $E$ {\it without} applying $\tilde W$, the ${\bar B}$ part of which can be brought forward to the very beginning of the whole procedure. Therefore, as presented by (\ref{eq:uiuiiw}), the state $(\Psi^{AC})^{\otimes n}$ is almost invariant by the action of the quantum operation ${\mathcal E}_2\circ{\mathcal E}_1:{\bar A}\rightarrow{\bar A}_c$, which is defined by (\ref{eq:kamenaref}) and is indicated by the gray shaded region in the figure.}
\label{fig:markovmeasureeee}
\end{figure}

Consider a direct-sum decomposition of ${\mathcal H}^{\bar A}\otimes{\mathcal H}^{\bar B}\otimes{\mathcal H}^{E}$ as
\begin{eqnarray}
{\mathcal H}^{\bar A}\otimes{\mathcal H}^{\bar B}\otimes{\mathcal H}^{E}={\mathcal H}_{{\mathcal R}_n}\oplus{\mathcal H}_{\perp},\nonumber
\end{eqnarray}
where ${\mathcal H}_{{\mathcal R}_n}$ is the support of $\Psi_{{\mathcal R}_n}^{{\bar A}{\bar B}E}$ and ${\mathcal H}_{\perp}$ is its orthogonal complement. Letting $I_\perp$ be the identity operator on ${\mathcal H}_{\perp}$, define a unitary isomorphism ${\tilde W}:{\mathcal H}^{\bar A}\otimes{\mathcal H}^{\bar B}\otimes{\mathcal H}^{E}\rightarrow{\mathcal H}^{\bar B}\oplus{\mathcal H}_{\perp}$ by ${\tilde W}:=W^\dagger\oplus I_\perp$. Equality (\ref{eq:ariariariari}) then implies
\begin{eqnarray}
&&\!\!\!\!\!\!\!\!\!\!\!\!\!\!\!\!\left\|{\tilde W}U_2U_1|\Psi^{\otimes n}\rangle\!\langle\Psi^{\otimes n}|^{{\bar A}{\bar B}{\bar C}}U_1^\dagger U_2^\dagger{\tilde W}^\dagger\right.\nonumber\\
&&\;\;\;\;\;\;\;\;\;\;\;\;\left.-|\Psi^{\otimes n}\rangle\!\langle\Psi^{\otimes n}|^{{\bar A}_c{\bar B}{\bar C}}\right\|_1\leq\epsilon.\label{eq:uiuiiww}
\end{eqnarray}
as depicted in Figure \ref{fig:markovmeasureeee}. Define linear CPTP maps ${\mathcal E}_1:{\bar A}\rightarrow G$ and ${\mathcal E}_2:G\rightarrow{\bar A}_c$ by
\begin{eqnarray}
{\mathcal E}_1(\cdot)={\rm Tr}_{\bar A}[U_1(\cdot)U_1^\dagger],\;\;{\mathcal E}_2(\cdot)={\rm Tr}_{E}[U_2(\cdot)U_2^\dagger].\label{eq:kamenaref}
\end{eqnarray}
By taking the partial trace in (\ref{eq:uiuiiww}) so that the remaining system is ${\bar A}_c{\bar C}$ (see Figure \ref{fig:markovmeasureeee}), we obtain
\begin{eqnarray}
\left\|({\mathcal E}_2\circ{\mathcal E}_1)((\Psi^{\otimes n})^{{\bar A}{\bar C}})-(\Psi^{\otimes n})^{{\bar A}_c{\bar C}}\right\|_1\leq\epsilon.\nonumber
\end{eqnarray}
Therefore, we see from Lemma \ref{thm:corrtran} and Inequality (\ref{eq:contmii}), that
\begin{eqnarray}
&&\frac{1}{n}I({\bar A}_c:{\bar B}{\bar C})_{({\mathcal E}_2\circ{\mathcal E}_1)(\Psi^{\otimes n})}\nonumber\\
&&\geq M_{A|B}^D(\Psi^{ABC})-5\eta(\zeta_{{}_\Psi}\!(\epsilon))\log{d_A},\nonumber
\end{eqnarray}
and also
\begin{eqnarray}
&&\frac{1}{n}I(G:{\bar B}{\bar C})_{{\mathcal E}_1(\Psi^{\otimes n})}\nonumber\\
&&\geq M_{A|B}^D(\Psi^{ABC})-5\eta(\zeta_{{}_\Psi}\!(\epsilon))\log{d_A}\label{eq:igeqm}
\end{eqnarray}
by the monotonicity of the quantum mutual information. 

From (\ref{eq:kakugohaiika}), (\ref{eq:kamenaref}) and (\ref{eq:igeqm}), we have
\begin{eqnarray}
nR&\geq&S(G)_{\Psi_{{\mathcal V}_n}}\nonumber\\
&=&I(G:{\bar B}{\bar C})_{\Psi_{{\mathcal V}_n}}-S({\bar B}{\bar C})_{\Psi_{{\mathcal V}_n}}+S(G{\bar B}{\bar C})_{\Psi_{{\mathcal V}_n}}\nonumber\\
&=&I(G:{\bar B}{\bar C})_{{\mathcal E}_1(\Psi^{\otimes n})}-S({\bar B}{\bar C})_{\Psi^{\otimes n}}+S({\bar A})_{\Psi_{{\mathcal V}_n}}\nonumber\\
&=&I(G:{\bar B}{\bar C})_{{\mathcal E}_1(\Psi^{\otimes n})}-S({\bar A})_{\Psi^{\otimes n}}+S({\bar A})_{{\mathcal V}_n(\Psi^{\otimes n})}\nonumber\\
&\geq&I(G:{\bar B}{\bar C})_{{\mathcal E}_1(\Psi^{\otimes n})}\nonumber\\
&\geq& nM_{A|B}^D(\Psi^{ABC})-5n\eta(\zeta_{{}_\Psi}\!(\epsilon))\log{d_A},\nonumber
\end{eqnarray}
where the second inequality follows from the monotonicity of the von Neumann entropy under random unitary operations. This completes the proof by taking the limit of $\epsilon\rightarrow0$ and noting Inequality (\ref{eq:leftmark}). \hfill$\blacksquare$

\subsection{Proof of Equality (\ref{eq:rightmarkpure}) Under the Additional Constraint (\ref{eq:additional})}

First we prove $M^{R}_{A|AB}(\Psi^{ABC})\leq M^D_{A|B}(\Psi^{ABC})$. By definition, for any $R>M_{A|B}^D(\Psi^{ABC})$, there exists a sequence of sets of unitaries $\{\{V_{n,k}\}_{k=1}^{2^{nR}}\}_{n=1}^\infty$, with each $V_{n,k}$ acting on $({\mathcal H}^A)^{\otimes n}$, such that ${\mathcal V}_n(\Psi^{\otimes n})$ is $\epsilon_n$-decomposable on ${\bar B}$ for ${\mathcal V}_n:\tau\mapsto2^{-nR}\sum_{k=1}^{2^{nR}}V_{n,k}\tau V_{n,k}^{\dagger}$ and $\lim_{n\rightarrow\infty}\epsilon_n=0$. From Lemma \ref{lmm:markovrecoverable}, it follows that ${\mathcal V}_n(\Psi^{\otimes n})$ is $2\epsilon_n$-recoverable from $AB$. In addition, it is proved in \cite{waka15_markov_paper} (see Appendix B-B therein) that $\epsilon_n$ vanishes exponentially with $n$, which implies
\begin{eqnarray}
\lim_{n\rightarrow\infty}n\cdot 2\epsilon_n=0.\nonumber
\end{eqnarray}
Since this relation holds for any $R>M_{A|B}^D(\Psi^{ABC})$, we obtain $M^{R}_{A|AB}(\Psi^{ABC})\leq M^D_{A|B}(\Psi^{ABC})$.

Second we prove $M^{R}_{A|AB}(\Psi^{ABC})\geq M^D_{A|B}(\Psi^{ABC})$. For any $R>M^R_{A|AB}(\Psi^{ABC})$, there exists a sequence of sets of unitaries $\{\{V_{n,k}\}_{k=1}^{2^{nR}}\}_{n=1}^\infty$ such that ${\mathcal V}_n(\Psi^{\otimes n})$ is $\epsilon_n$-recoverable from ${\bar A}{\bar B}$ for ${\mathcal V}_n:\tau\mapsto2^{-nR}\sum_{k=1}^{2^{nR}}V_{n,k}\tau V_{n,k}^{\dagger}$ and $\lim_{n\rightarrow\infty}n\cdot\epsilon_n=0$. From Lemma \ref{lmm:markovrecoverable}, it follows that ${\mathcal V}_n((\Psi^{ABC})^{\otimes n})$ is $f(\epsilon_n,d_C^n)$-recoverable from ${\bar B}{\bar C}$. Note that we have
\begin{eqnarray}
f(\epsilon_n,d_C^n)&=&2\sqrt{\ln{2}}\sqrt{4\epsilon_n\log{d_C^n}+2h(\epsilon_n)}\nonumber\\
&=&2\sqrt{\ln{2}}\sqrt{4n\epsilon_n\log{d_C}+2h(\epsilon_n)},\nonumber
\end{eqnarray}
which implies $\lim_{n\rightarrow\infty}f(\epsilon_n,d_C^n)=0$. Since this relation holds for any $R>M_{A|AB}^R(\Psi^{ABC})$, we obtain $M_{A|AB}^R(\Psi^{ABC})\geq M_{A|BC}^R(\Psi^{ABC})=M_{A|B}^D(\Psi^{ABC})$.\hfill$\blacksquare$

\section{Proof of Theorem \ref{thm:rightmark}}\label{app:measmark}

First we prove $M^{R,m}_{A|AB}(\Psi^{ABC})\leq M^D_{A|B}(\Psi^{ABC})$. From Theorem \ref{thm:leftmarkpure}, for any $R>M_{A|AB}^R(\Psi^{ABC})$, there exists a sequence of sets of unitaries $\{\{V_{n,k}\}_{k=1}^{2^{nR}}\}_{n=1}^{\infty}$, with each $V_{n,k}$ acting on $({\mathcal H}^A)^{\otimes n}$, such that ${\mathcal V}_n(\Psi^{\otimes n})$ is $\epsilon_n$-recoverable from ${\bar A}{\bar B}$ for ${\mathcal V}_n:\tau\mapsto2^{-nR}\sum_{k=1}^{2^{nR}}V_{n,k}\tau V_{n,k}^{\dagger}$ and $\lim_{n\rightarrow\infty}n\cdot\epsilon_n=0$. For each $n$, let $\ket{{\varrho}}^{A_0G}$ be a maximally entangled state with Schmidt rank $2^{nR}$. Using $\{V_{n,k}\}_{k=1}^{2^{nR}}$, construct Alice's measurement ${{\mathbb M}}=\{M^{{\bar A}A_0\rightarrow{\bar A}}_{ k}\}_{k=1}^{2^{nR}}$ as
\begin{eqnarray}
M^{{\bar A}A_0\rightarrow{\bar A}}_{ k}=\frac{1}{\sqrt{2^{nR}}}\sum_{j=1}^{2^{nR}}\exp{\left(i\frac{2\pi jk}{2^{nR}}\right)}\langle j|^{A_0}\otimes V_{n,j}^{\bar A}\nonumber\\
(k\in[1,2^{nR}]).\nonumber
\end{eqnarray}
It is straightforward to verify that, for any $k$, the post-measurement state $\Psi_{ k}$ satisfies $\Psi^{{\bar A}{\bar B}{\bar C}}_{ k}={\mathcal V}_n(\Psi^{\otimes n})$, which implies that Condition 2) is satisfied by the correspondence $\epsilon\rightarrow\epsilon_n$. Condition 1) is met as well, since random unitary operations on $\bar A$ does not change the reduced state on ${\bar B}{\bar C}$ at all. Condition 3) is also satisfied, because we have 
\begin{eqnarray}
I(G:{\bar B}{\bar C})_{\Psi_k}&=&S(G)_{\Psi_k}+S({\bar B}{\bar C})_{\Psi_k}-S({\bar B}{\bar C}G)_{\Psi_k}\nonumber\\
&=&S(G)_{\Psi_k}+nS(BC)_{\Psi}-S({\bar A})_{\Psi_k}\nonumber\\
&\leq&S(G)_{\Psi_k}+nS(BC)_{\Psi}-nS(A)_{\Psi}\nonumber\\
&=&S(G)_{\Psi_k}\leq nR,\nonumber
\end{eqnarray}
where the third line follows by the monotonicity of the von Neumann entropy under random unitary operations. Thus, a pair $(|\varrho\rangle,{\mathbb M})$ is an $(n,R,\epsilon_n)$-Markovianization pair for $|\Psi\rangle^{ABC}$. Since this relation holds for any $R>M_{A|AB}^R(\Psi^{ABC})$ and each $n$, we obtain $M^{R,m}_{A|AB}(\Psi^{ABC})\leq M^R_{A|AB}(\Psi^{ABC})=M^D_{A|B}(\Psi^{ABC})$.

Second we prove $M^{R,m}_{A|AB}(\Psi^{ABC})\geq M^D_{A|B}(\Psi^{ABC})$. Fix arbitrary $n\in{\mathbb N}$ and $\epsilon>0$ satisfying
\begin{eqnarray}
16\epsilon<n\leq\frac{1}{4\epsilon},\label{eq:steps}
\end{eqnarray}
and consider a measurement $\{M^{{\bar A}A_0\rightarrow A'}_{ k}\}_k$ and a state $\ket{{\varrho}}^{A_0G}$ that satisfy Inequalities (\ref{eq:recoverabilite2}) and (\ref{eq:recoverabilize}). Define
\begin{eqnarray}
&&\delta_k:=\left\|(\Psi^{\otimes n})^{{\bar B}{\bar C}}-\Psi^{{\bar B}{\bar C}}_{ k}\right\|_1,\label{eq:epsilonk}\\
&&\delta_k':=\left\|\Psi^{A'{\bar B}{\bar C}}_{ k}-{\mathcal R}_{ k}(\Psi^{A'{\bar B}}_{ k})\right\|_1.\label{eq:epsilonprimek}
\end{eqnarray}
Fix one $k$ for the moment, and assume $\delta_k<1/4$, $\delta_k'<1$.  Lemma \ref{lmm:markovrecoverable} and (\ref{eq:epsilonprimek}) imply there exist quantum operations ${\mathcal R}'_{ k}:{\bar B}\rightarrow A'{\bar B}$ such that
\begin{eqnarray}
\left\|\Psi^{A'{\bar B}{\bar C}}_{ k}-{\mathcal R}'_{ k}(\Psi^{{\bar B}{\bar C}}_{ k})\right\|_1\leq f(\delta_k',d_C^n).\nonumber
\end{eqnarray}
From (\ref{eq:epsilonk}) and the monotonicity of the trace distance, we have
\begin{eqnarray}
\left\|{\mathcal R}'_{ k}((\Psi^{\otimes n})^{{\bar B}{\bar C}})-{\mathcal R}'_{ k}(\Psi^{{\bar B}{\bar C}}_{ k})\right\|_1\leq\delta_k.\nonumber
\end{eqnarray}
By the triangle inequality, we obtain
\begin{eqnarray}
\left\|\Psi^{A'{\bar B}{\bar C}}_{ k}-{\mathcal R}'_{ k}((\Psi^{\otimes n})^{{\bar B}{\bar C}})\right\|_1\leq \delta_k+f(\delta_k',d_C^n).
\label{eq:errorsk}
\end{eqnarray}
Let $W:{\bar B}\rightarrow{A'}{\bar B}E$ be an isometry such that a Stinespring dilation of ${\mathcal R}'_{ k}$ is given by ${\mathcal R}'_{ k}(\tau)={\rm Tr}_E[W\tau W^\dagger]$. Then a purification of ${\mathcal R}'_{ k}((\Psi^{\otimes n})^{{\bar B}{\bar C}})$ is given by
\begin{eqnarray}
|\Psi_{W}\rangle^{{A'}{\bar B}{\bar C}{\bar {A}}E}:=W|\Psi^{\otimes n}\rangle^{{\bar {A}}{\bar B}{\bar C}}.\label{eq:wk}
\end{eqnarray} 
Due to (\ref{eq:errorsk}) and Uhlmann's theorem\cite{uhlmann76}, there exists an isometry $U_2:G\rightarrow{\bar A}E$ such that
\begin{eqnarray}
\left\|U_2|\Psi_{ k}\rangle\!\langle\Psi_{k}|U_2^\dagger-|\Psi_{W}\rangle\!\langle\Psi_{W}|\right\|_1\leq2\sqrt{\delta_k+f(\delta_k',d_C^n)}.\label{eq:utwok}
\end{eqnarray}
On the other hand, (\ref{eq:epsilonk}) implies there exists another isometry $U_1:{\bar A}\rightarrow A'G$ such that
\begin{eqnarray}
\left\|U_1|\Psi\rangle\!\langle\Psi|^{\otimes n}U_1^\dagger-\proj{\Psi_{ k}}^{A'{\bar B}{\bar C}G}\right\|_1\leq2\sqrt{\delta_k}.
\label{eq:psikchi}
\end{eqnarray}
From (\ref{eq:wk}), (\ref{eq:utwok}) and (\ref{eq:psikchi}), we obtain
\begin{eqnarray}
&&\!\!\!\!\!\!\!\!\!\!\!\!\!\!\!\!\left\|U_2U_1|\Psi^{\otimes n}\rangle\!\langle\Psi^{\otimes n}|^{{\bar A}{\bar B}{\bar C}}U_1^\dagger U_2^\dagger-W|\Psi^{\otimes n}\rangle\!\langle\Psi^{\otimes n}|^{{\bar A}{\bar B}{\bar C}}W^\dagger\right\|_1\nonumber\\
&&\leq2\sqrt{\delta_k}+2\sqrt{\delta_k+f(\delta_k',d_C^n)}.\label{eq:uiuiiw}
\end{eqnarray} 

Define linear CPTP maps ${\mathcal E}_1:{\bar A}\rightarrow G$ and ${\mathcal E}_2:G\rightarrow{\bar A}$ by
\begin{eqnarray}
{\mathcal E}_1(\cdot)={\rm Tr}_{A'}[U_1(\cdot)U_1^\dagger],\;\;{\mathcal E}_2(\cdot)={\rm Tr}_{E}[U_2(\cdot)U_2^\dagger].\nonumber
\end{eqnarray}
By tracing out $A'$ in (\ref{eq:psikchi}), we have
\begin{eqnarray}
\left\|{\mathcal E_1}(|\Psi\rangle\!\langle\Psi|^{\otimes n})-\Psi^{{\bar B}{\bar C}G}_{ k}\right\|_1\leq2\sqrt{\delta_k}.\nonumber
\end{eqnarray}
Thus we obtain
\begin{eqnarray}
&&I(G:{\bar B}{\bar C})_{\Psi_{ k}}-nM^D_{A|B}(\Psi^{ABC})\nonumber\\
&\geq&I(G:{\bar B}{\bar C})_{{\mathcal E}_1(\Psi^{\otimes n})}-nM^D_{A|B}(\Psi^{ABC})\nonumber\\
&&-5n\eta(2\sqrt{\delta_k})\log{(d_Bd_C)}\nonumber
\end{eqnarray}
by Inequality (\ref{eq:contmii}). The same method as used to obtain (\ref{eq:igeqm}) from (\ref{eq:uiuiiww}), also shows
\begin{eqnarray}
&&\!\!\!\!\!\!\!\!\!\!\!\!\!\!\!I(G:{\bar B}{\bar C})_{{\mathcal E}_1(\Psi^{\otimes n})}-nM_{A|B}(\Psi^{ABC})\nonumber\\
&&\!\!\!\!\!\!\!\!\!\!\!\!\!\!\!\geq-5(\eta\circ\zeta_\Psi)\!\left(2\sqrt{\delta_k}+2\sqrt{f(\delta_k',d_C^n)}\right)\log{(d_Bd_C)}\nonumber
\end{eqnarray}
from (\ref{eq:uiuiiw}) due to Lemma \ref{thm:corrtran}. Here, we denote a function $\eta(\zeta_\Psi(\cdot))$ by $(\eta\circ\zeta_\Psi)(\cdot)$. Thus we obtain
\begin{eqnarray}
I(G:{\bar B}{\bar C})_{\Psi_{ k}}\!\!-\!nM^D_{A|B}(\Psi^{ABC})\geq-n\xi_k\log{(d_Bd_C)},\label{eq:ariari}
\end{eqnarray}
where we defined
\begin{eqnarray}
\xi_{k}:=5\eta(2\sqrt{\delta_k})+5(\eta\circ\zeta_\Psi)\!\left(2\sqrt{\delta_k}+2\sqrt{f(\delta_k',d_C^n)}\right)\nonumber
\end{eqnarray}
for $k$ such that $\delta_k<1/4$ and $\delta_k'<1$.

Without loss of generality, we assume that ${\rm supp}\:\Psi^A={\mathcal H}^A$. Since $\Psi^{ABC}$ is a pure state, we see that
\begin{eqnarray}
M^D_{A|B}(\Psi^{ABC})\leq2\log{d_A}\leq2\log{(d_Bd_C)}.\label{eq:mudamuda}
\end{eqnarray}
Consider an arbitrary $k\in{\mathbb K}$ and define
\begin{eqnarray}
\xi_k':=\begin{cases}
\min\{\xi_k,2\}&\text{if $\delta_k<1/4$ and $\delta_k'<1$}\\
2&\text{otherwise}
\end{cases}.\label{eq:hurueruzo}
\end{eqnarray}
Combining (\ref{eq:ariari}) and (\ref{eq:mudamuda}),
\begin{eqnarray}
I(G:{\bar B}{\bar C})_{\Psi_{ k}}-nM^D_{A|B}(\Psi^{ABC})\geq-n\xi_k\log{(d_Bd_C)}\nonumber
\end{eqnarray}
for any $k\in{\mathbb K}$. Thus we obtain
\begin{eqnarray}
I(G:{\bar B}{\bar C})_{av}\geq nM^D_{A|B}(\Psi^{ABC})-n\xi_{av}^{({ n,\epsilon})}\log{(d_Bd_C)},\label{eq:ibcgave}
\end{eqnarray}
where $\xi_{av}^{({ n,\epsilon})}:=\sum_kp_{k}\xi_k$.

Let us now evaluate $\xi_{av}^{({ n,\epsilon})}$. For any $\lambda>0$, define two sets ${\mathbb K}_{\rm inv}(\lambda)\in{\mathbb K}$ and ${\mathbb K}_{\rm rec}(\lambda)\in{\mathbb K}$ by
\begin{eqnarray}
&&{\mathbb K}_{\rm inv}(\lambda):=\left\{k\in{\mathbb K}\left|\|(\Psi^{\otimes n})^{{\bar B}{\bar C}}-\Psi^{{\bar B}{\bar C}}_{ k}\|_1\leq\lambda\right.\right\},\nonumber\\
&&{\mathbb K}_{\rm rec}(\lambda):=\left\{k\in{\mathbb K}\left|\|\Psi^{A'{\bar B}{\bar C}}_{ k}-{\mathcal R}_{ k}(\Psi^{A'{\bar B}}_{ k})\|_1\leq\lambda\right.\right\}.\nonumber
\end{eqnarray}
From Conditions (\ref{eq:recoverabilite2}) and (\ref{eq:recoverabilize}), for any $t\geq1$ we have
\begin{eqnarray}
\sum_{k\notin {\mathbb K}_{\rm inv}(t{\epsilon})}p_{k}&=&\frac{1}{t{\epsilon}}\sum_{k\notin {\mathbb K}_{\rm inv}(t{\epsilon})}p_{k}t{\epsilon}\;\;\leq\;\;\frac{1}{t{\epsilon}}\sum_{k\notin {\mathbb K}_{\rm inv}(t{\epsilon})}p_{k}\delta_k\nonumber\\
&\leq&\frac{1}{t{\epsilon}}\sum_{k\in {\mathbb K}}p_{k}\delta_k\leq\frac{1}{t},\nonumber
\end{eqnarray}
and similarly, have
\begin{eqnarray}
\sum_{k\notin{\mathbb K}_{\rm rec}(t{\epsilon})}p_{k}\leq\frac{1}{t},\nonumber
\end{eqnarray}
which leads to
\begin{eqnarray}
\sum_{k\notin{\mathbb K}_{\rm inv}(t\epsilon)\cap{\mathbb K}_{\rm rec}(t\epsilon)}p_{k}\leq\sum_{k\notin {\mathbb K}_{\rm inv}(t\epsilon)}p_{k}+\sum_{k\notin{\mathbb K}_{\rm rec}(t\epsilon)}p_{k}\leq\frac{2}{t}.\nonumber
\end{eqnarray}
Due to (\ref{eq:hurueruzo}), this shows that
\begin{align}
\xi_{av}^{({ n,\epsilon})}&=\sum_{k\in{\mathbb K}_{\rm inv}(t{\epsilon})\cap{\mathbb K}_{\rm rec}(t{\epsilon})}p_{k}\xi_k'+\sum_{k\notin{\mathbb K}_{\rm inv}(t{\epsilon})\cap{\mathbb K}_{\rm rec}(t{\epsilon})}p_{k}\xi_k'\nonumber\\
&\leq5\eta(2\sqrt{t{\epsilon}})+{ 5(\eta\circ\zeta_\Psi)}\!\left(2\sqrt{t{\epsilon}}+2\sqrt{f(t{\epsilon},d_C^n)}\right)+\frac{4}{t}\nonumber\\\label{eq:rockman}
\end{align}
when $t{\epsilon}<1/4$. Let $t=1/\sqrt{n{\epsilon}}$. Noting that $\sqrt{n\epsilon}\leq1/2$, $t\geq2$ and $t\epsilon<1/4$ from (\ref{eq:steps}), and that
\begin{eqnarray}
t{\epsilon}\leq\sqrt{\epsilon},\;\;\;f(t{\epsilon},d_C^n)\leq f(\sqrt{n{\epsilon}},d_C),\nonumber
\end{eqnarray}
we have
\begin{align}
\xi_{av}^{(n,\epsilon)}&\leq5\eta(2\sqrt[4]{\epsilon})+5(\eta\circ\zeta_\Psi)\!\left(2\sqrt[4]{\epsilon}+2\sqrt{f(\sqrt{n{\epsilon}},d_C)}\right)\nonumber\\
&\quad\quad+4\sqrt{n{\epsilon}}\nonumber\\
&\leq\xi(n{\epsilon}),\label{eq:xineps}
\end{align}
where we defined a function 
\begin{align}
\xi(x)&:=5\eta(2\sqrt[4]{x})+5(\eta\circ\zeta_\Psi)\!\left(2\sqrt[4]{x}+2\sqrt{f(\sqrt{x},d_C)}\right)\nonumber\\
&\quad\quad+4\sqrt{x}.\label{eq:dddxix}
\end{align}
Substituting (\ref{eq:xineps}) to (\ref{eq:ibcgave}) yields
\begin{eqnarray}
I(G:{\bar B}{\bar C})_{av}\geq nM^D_{A|B}(\Psi^{ABC})-n\xi(n\epsilon)\log{(d_Bd_C)}.\label{eq:ibcgaveee}
\end{eqnarray}

Suppose now that $R>M^{R,m}_{A|AB}(\Psi^{ABC})$. By definition, there exists a sequence of $(n,R,\epsilon_n)$-Markovianization pairs for $|\Psi\rangle^{ABC}$ such that $\lim_{n\rightarrow\infty}\epsilon_n=0$. From Inequalities (\ref{eq:iavebou}) and (\ref{eq:ibcgaveee}), we have
\begin{eqnarray}
R\geq M^D_{A|B}(\Psi^{ABC})-\xi(n\cdot\epsilon_n)\log{(d_Bd_C)}
\end{eqnarray}
for each $n$, which leads to $R\geq M^D_{A|B}(\Psi^{ABC})$ because of Condition (\ref{eq:additionall}). Note that we have
\begin{eqnarray}
\lim_{x\rightarrow0}\xi(x)=0\label{eq:convxix}
\end{eqnarray}
from (\ref{eq:dddxix}). Since this relation holds for any $R>M^{R,m}_{A|AB}(\Psi^{ABC})$, we obtain $M^{R,m}_{A|AB}(\Psi^{ABC})\geq M^D_{A|B}(\Psi^{ABC})$. \hfill$\blacksquare$

\begin{rmk}
If Proposition \ref{prp:symrec} is true, $f(t{\epsilon},d_C^n)$ in (\ref{eq:rockman}) is replaced by $g(t{\epsilon})$. Substituting $1/\sqrt{{\epsilon}}$ to $t$ and $\epsilon_n$ to $\epsilon$, it follows that $\lim_{n\rightarrow\infty}\xi_{av}^{(n,\epsilon_n)}=0$ if only $\lim_{n\rightarrow\infty}\epsilon_n=0$. Thus we have $M^{R,m}_{A|AB}(\Psi^{ABC})\geq M^D_{A|B}(\Psi^{ABC})$ irrespective of Condition (\ref{eq:additionall}).
\end{rmk}



%

\end{document}